\documentclass[12pt]{article}
\usepackage[colorlinks,linkcolor=Blue,citecolor=Blue,bookmarks,bookmarksnumbered]{hyperref}
\usepackage{charter}
\usepackage[scaled=0.85]{helvet}
\usepackage{XXXE,accents,mathtools}
\usepackage{graphicx,color}
\usepackage{booktabs,array}
\usepackage{multirow}
\usepackage{placeins}

\def\Dt#1{\accentset{\hbox{\large.}}{#1}}	
\def\DDt#1{\accentset{\hbox{\large.\kern-2pt.}}{#1}}	
\def\dt#1{\accentset{\hbox{\normalsize.}}{#1}}	
\def\ddt#1{\accentset{\hbox{\large\kern.5pt.\kern-1pt.}}{#1}}	
\def\cb#1#2{\setlength\fboxsep{1pt}\colorbox{#1}{\color{#1}\fbox{\color{black}#2}}}
\def\cB#1{\hbox to0pt{\setlength\fboxsep{0pt}\hss\color{grey3}\fbox{\cb{white}{#1}}\hss}}
\def\bB#1{\hbox to0pt{\setlength\fboxsep{0pt}\hss\color{grey3}\fbox{\cb{black}{\color{white}#1}}\hss}}

\def\eX{\rlap{\raisebox{2pt}{\kern1.125pt\scriptsize\it=}}{X}}
\def\eY{\rlap{\raisebox{1pt}{\kern-0.125pt\scriptsize\textsl{=}}}{Y}}
\def\EY{\rlap{\raisebox{.875pt}{\kern.625pt\scriptsize\textsl{=}}}{\cal Y}}
\def\feY{\rlap{\raisebox{1.5pt}{\kern.5pt\tiny\it=}}{Y}}
\def\eU{\rlap{\raisebox{1.5pt}{\kern1.25pt\scriptsize\it=}}{\Y}}
\def\feU{\rlap{\raisebox{1.5pt}{\kern.95pt\tiny\it=}}{\Y}}

\let\bs=\boldsymbol
\def\rD{{\rm D}}

\def\bDb{\hbox{\kern2pt\vrule height10pt depth-9.2pt width6pt\kern-9pt{$\boldsymbol D$}}}

\def\bQb{\hbox{\kern2pt\vrule height10pt depth-9.2pt width6pt\kern-9pt{$\boldsymbol Q$}}}

\def\BSb{{\overline{\boldsymbol\Sigma}}}
\def\BSb{\hbox{\kern.5pt\vrule height10pt depth-9.2pt width6pt\kern-7.5pt{$\boldsymbol\S$}}}

\def\rQb{\hbox{\kern1pt\vrule height10pt depth-9.2pt width6pt\kern-8pt{\bf Q}}}

\def\vC#1{\vcenter{\hbox{\hss#1\hss}}}

\def\rBx#1#2{\hbox to#1{#2\hss}}

\def\cY{{\cal Y}}
\def\vdt{\partial_\tau}

\definecolor{Hey}{rgb}{.9,.05,.4}
\definecolor{plum}{rgb}{.4,0,.6}

\definecolor{Green}  {rgb}{0.10,0.70,0.10} 
\definecolor{Orange} {rgb}{1.00,0.50,0.15} 
\definecolor{Red}    {rgb}{0.90,0.00,0.12} 
\definecolor{Purple} {rgb}{0.42,0.15,0.45} 
\definecolor{Turque} {rgb}{0.00,0.65,0.85} 
\definecolor{Blue}   {rgb}{0.00,0.00,1.00} 
\definecolor{Magenta}{rgb}{1.00,0.00,1.00} 
\definecolor{Gold}   {rgb}{1.00,0.75,0.25} 
\definecolor{Seaweed}{rgb}{0.01,0.24,0.09} 
\definecolor{Brown}  {rgb}{0.43,0.26,0.32} 
\definecolor{grey1}  {rgb}{0.20,0.20,0.20} 
\definecolor{grey2}  {rgb}{0.40,0.40,0.40} 
\definecolor{grey3}  {rgb}{0.60,0.60,0.60} 
\definecolor{grey4}  {rgb}{0.80,0.80,0.80} 
\definecolor{grey5}  {rgb}{0.90,0.90,0.90} 
\def\C#1#2{{\ifcase#1\or
             \color{Green}\or \color{Orange}\or \color{Red}\or
              \color{Purple}\or \color{Turque}\or \color{Blue}\or
               \color{Magenta}\or \color{Gold}\or \color{Seaweed}\or
                \color{Brown}\or\color{grey1}\or\color{grey2}\or
                 \color{grey3}\else\color{grey4}\fi#2}}

 \SfTitles 
 \allowdisplaybreaks
\begin{document}
\thispagestyle{empty}
 \noindent
 \today\hfill
  \vspace*{5mm}
 \begin{center}
{\LARGE\sf\bfseries On the Construction and the Structure\\[1mm]
                    of Off-Shell Supermultiplet Quotients}\\[1mm]
  \vspace*{5mm}
 \begin{tabular}{p{80mm}cp{80mm}}
 \hfill
{\large\sf\bfseries  Tristan H\"{u}bsch$^{*\dag}$}
                    &and&
{\large\sf\bfseries  Gregory A.~Katona$^{\dag\ddag}$}\\[1mm]
\MC3c{\small\it
  $^*$\,Department of Physics \&\ Astronomy,
  Howard University, Washington, DC 20059} \\[-1mm]
\MC3c{\small\it
  $^\dag$\,Department of Physics, University
  of Central Florida, Orlando, FL 32816}\\[-1mm]
\MC3c{\small\it
  $^\ddag$\,Affine Connections, LLC, College Park, MD 20740}\\[0mm]
 \hfill {\tt  thubsch@howard.edu}&&{\tt grgktn@knights.ucf.edu}
  \end{tabular}\\[1mm]
  \vspace*{5mm}
{\sf\bfseries ABSTRACT}\\[2mm]
\parbox{151mm}{Recent efforts to classify representations of supersymmetry with no central charge\cite{r6-3.1} have focused on supermultiplets that are aptly depicted by Adinkras, wherein every supersymmetry generator transforms each component field into precisely one other component field or its derivative.
Herein, we study gauge-quotients of direct sums of Adinkras by a supersymmetric image of another Adinkra and thus solve a puzzle from Ref.\cite{r6-1}: Such (gauge-)quotients are not Adinkras but more general types of supermultiplets, each depicted as a connected network of Adinkras.
 Iterating this gauge-quotient construction then yields an indefinite sequence of ever larger supermultiplets, reminiscent of Weyl's construction that is known to produce all finite-dimensional unitary representations in Lie algebras.
 } 
\end{center}
\vspace{5mm}
\noindent
\parbox[t]{60mm}{PACS: {\tt11.30.Pb}, {\tt12.60.Jv}}\hfill
\parbox[t]{100mm}{\raggedleft\small\baselineskip=12pt\sl
            Program construction consists of\,\\[-1pt]
            a sequence of refinement steps.\\[-0pt]
            |~Niklaus Wirth}

\section{Introduction, Results and Summary}
 \label{IRS}
Recent efforts to classify off-shell representations of $N$-extended worldline supersymmetry with no central charge%
\cite[ and references therein]{r6-3.1}
 developed a detailed classification of a huge class (${\sim}\,10^{12}$ for $N\leq32$) of such supermultiplets, wherein each supercharge maps each component field to precisely one other component field or its derivative, and which are faithfully represented by graphs called {\em\/Adinkras\/}; see also Refs.\cite{rPT,rT01,rT01a,rCRT,rKRT,rKT07,rGKT10} for related work. Refs.\cite{r6-1,r6-1.2} also provide rigorous constructive theorems that relate Adinkras to standard superfield expressions, allowing us to toggle easily and without much special notice between these representations.

Our motivation for focusing on worldline supersymmetry is threefold:
 ({\bf1})~Supersymmetry in all higher-dimensional spacetime always contains worldline supersymmetry by way of dimensional reduction of spacetime to (proper or some coordinate) time.
 ({\bf2})~The {\em\/underlying\/} theoretical framework for the description of the dynamics of extended objects such as $M$-theory includes worldline supersymmetry in a prominent way.
 Finally,
 ({\bf3})~the Hilbert space of any supersymmetric field theory necessarily admits the action of an induced worldline supersymmetry.
 Any one of these theories would require off-shell fields for a self-consistently quantum formulation, perhaps using Feynman-Hibbs path integrals, which then provides our fundamental motivation for exploring the structure of off-shell representations of $N$-extended worldline supersymmetry.

Unlike the well-developed representation theory of Lie algebras\cite{rWyb,rHall} and on-shell representation theory of supersymmetry\cite{rYM97Gau,rDF,rVSV}, classification and constructive algorithms for off-shell supermultiplets are far from complete. In fact the basic techniques of standard representation theory are incompatible with off-shell study: In all standard representation theory, representations are eigenspaces of mutually commuting (even) elements of the algebra. Although the Hamiltonian is a central element of the simplest type of supersymmetry algebra\eq{e:SuSy}, which is contained in all others, off-shell representations are expressly not eigenspaces of the Hamiltonian. To wit, any such eigenspace statement would constitute a spacetime differential equation, derivable as an Euler-Lagrange equation, and so make the eigenspace an on-shell representation.

Herein, we solve a puzzle raised in Ref.\cite{r6-1}, and show that this specifies a construction of indefinitely many and ever larger supermultiplets of $N$-extended worldline supersymmetry without central charges, for all $N\geq3$, using adinkraic (see below) off-shell supermultiplets as building blocks.
 Conceptually, the construction is not unlike the Weyl construction of all finite-dimensional unitary representations of classic Lie algebras (by tensoring the fundamental representation and then projecting through Young symmetrizations), but relies instead solely on supersymmetric reductions of direct {\em\/sums\/} of supermultiplets---something not possible with representations of Lie algebras. We focus on $N\,{=}\,3$ worldline supersymmetry for simplicity, and to disentangle the basic features of the construction from the complications stemming from additional symmetries (gauge and Lorentz) and structures (complex and hyper-complex). Nevertheless, our main result---see concluding section~\ref{s:Coda} and especially table~\ref{t:BoxDinkra} for a quick summary---will have straightforward generalizations to off-shell supermultiplets of all $(N\geqslant3)$-extended worldline supersymmetry.

In the remainder of this introduction, we specify the notation and conventions.
 Section~\ref{s:GQS} then presents a detailed analysis of the quotient supermultiplet, representable in superfield notation as $\IY_I/(i\rD_I\IX)$.
 In section~\ref{s:ZZ}, revisiting the indefinite sequence of supermultiplets constructed from Adinkras\cite{r6-1} in which $\IY_I/(i\rD_I\IX)$ is the first nontrivial example allows formulating our main result.
 Finally, section~\ref{s:Coda} contains our summary of prospects and conclusions, while two appendices collect some technical details deferred from the main part.

\paragraph{Worldline supersymmetry:}
As the simplest and (as it turns out) for our purposes nontrivial case, we focus on worldline $(N\,{=}\,3)$-supersymmetry algebra
\begin{equation}
 \left.
 \begin{aligned}
 \big\{\, Q_I \,,\, Q_J \,\big\}&=2\d_{IJ}\,H,&
 \big[\, H \,,\, Q_I \,\big] &=0,\\
 Q_I^{~\dagger}&= Q_I,& H^\dagger&=H,\quad
\end{aligned}\right\}\quad
 I,J=1,2,3,
  \label{e:SuSy}
\end{equation}
were $H=i\vdt$ is the worldline Hamiltonian.
 We also have the superderivatives\footnote{While not necessary, superspace methods incur no loss of generality\cite{rHTSSp08} and we find them convenient.}:
\begin{equation}
 \left.
 \begin{aligned}
 \big\{\, \rD_I \,,\, \rD_J \,\big\}&=2\d_{IJ}\,H,&
 \big[\, H \,,\, \rD_I \,\big] &=0,\\
 \rD_I^{~\dagger}&=-\rD_I,&
 \big\{\, Q_I \,,\, \rD_J \,\big\}&=0,\quad
\end{aligned}\right\}\quad
 I,J=1,2,3,
  \label{e:SuSyD}
\end{equation}
and the relationships:
\begin{equation}
  Q_I=i\rD_I-2i\d_{IJ}\q^J\,H,\quad\text{and}\quad
  \rD_I=-iQ_I-2\d_{IJ}\q^J\,H,
 \label{e:Q=D}
\end{equation}
where $\q^I$ provide the fermionic extension to (space)time into superspace. Note, however, that when applied on superfields (general functions over superspace), the $\rD_I$ act as left-derivatives while the $Q_I$ act as right-derivatives.
 Superfields are then used to represent supermultiplets, and component fields may then be defined as the $\q^I\to0$ evaluation of appropriately chosen superderivative superfield expression\cite{r1001}. In fact, supersymmetry transformation rules for the individual component fields may also be expressed using the $\rD_I$'s:
\begin{equation}
   Q_I(b) \id  i\rD_I\,\IB| \qquad\text{and}\qquad
   Q_I(f) \id -i\rD_I\,\IF|,
 \label{e:Q=DFJ}
\end{equation}
where
 $\IB$ and $\IF$ are appropriate superderivative superfield expressions defining the bosonic and fermionic differential expressions
 $b\Defl\IB|$ and $f\Defl\IF|$, respectively.

Then, an off-shell worldline supermultiplet is a collection of bosons and fermions
\begin{equation}
  \big(\f_1(\t),\dots,\f_m(\t)\,\big|
      \,\j_1(\t),\dots,\j_{m'}(\t)\big),
  \qquad m=m',
\end{equation}
such that
 each supercharge $Q_I$ transforms
 each boson into a linear combination of fermions and/or their $\t$-derivatives and {\em\/vice versa\/},
 and none of $(\f_i(\t)|\j_j(\t))$ are required to satisfy any differential equation that could be derived from some action as its classical equation of motion.

\paragraph{Adinkras:}%
The classification of Refs.~\cite{rA,r6-1,r6--1,r6-3c,r6-3,r6-3.2,r6-1.2,r6-3.4,r6-3.1} then focuses on supermultiplets wherein each $Q_I$ transforms every component field into precisely one other component field or its $\t$-derivative\ft{The worldline dimensional reductions of the best-known supermultiplets are of this form, such as the chiral and (real) vector supermultiplets of $N=1$ supersymmetry in $d=4$ spacetime\cite{r1001,rWB,rBK} and the twisted-chiral supermultiplets in $d=2$\cite{rTwSJG0,rGHR,rPhases,rTwSJG1,rMP1}.}.
 The $Q$-action is then unambiguously depicted by the graphs called ``Adinkras,'' following the ``dictionary'' provided in table~\ref{t:A}~\cite{r6-3}.
\begin{table}[ht]
  \centering
  \caption{\baselineskip=12pt Adinkras assign:
    (white/black) vertices to (boson/fermion) component fields;
    $I^\text{th}$ edge color to $Q_I$;
    solid/dashed edge to $\pm1$ signs in the right-hand side of the tabulated supersymmetry action;
    nodes are drawn at heights equal to the physical units of the depicted component (super)field.}
\vspace{2mm}
  \begin{tabular}{@{} cc|cc @{}}
    \makebox[15mm]{\bsf\boldmath Adinkra}
  & \makebox[40mm]{\bsf\boldmath $Q$-Action} 
  & \makebox[15mm]{\bsf\boldmath Adinkra}
  & \makebox[40mm]{\bsf\boldmath $Q$-Action} \\ 
    \hline
    \begin{picture}(5,9)(-1,5)
     \put(0,0){\includegraphics[height=11mm]{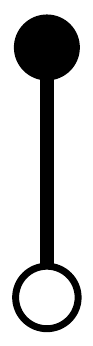}}
     \put(3,0){\scriptsize$k$}
     \put(3,9){\scriptsize$j$}
     \put(-1,4.5){\scriptsize$I$}
    \end{picture}\vrule depth4mm width0mm
     & $Q_I\begin{bmatrix}\j_j\\ \f_k\end{bmatrix}
           =\begin{bmatrix}i\Dt\f_k\\ \j_j\end{bmatrix}$
  & \begin{picture}(5,9)(-1,5)
     \put(0,0){\includegraphics[height=11mm]{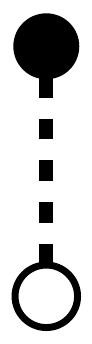}}
     \put(3,0){\scriptsize$k$}
     \put(3,9){\scriptsize$j$}
     \put(-1,4.5){\scriptsize$I$}
    \end{picture}\vrule depth4mm width0mm
     & $Q_I\begin{bmatrix}\j_j\\ \f_k\end{bmatrix}
           =\begin{bmatrix}-i\Dt\f_k\\-\j_j\end{bmatrix}$ \\[5mm]
    \hline
    \begin{picture}(5,9)(-1,5)
     \put(0,0){\includegraphics[height=11mm]{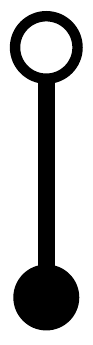}}
     \put(3,0){\scriptsize$j$}
     \put(3,9){\scriptsize$k$}
     \put(-1,4.5){\scriptsize$I$}
    \end{picture}\vrule depth4mm width0mm
     &  $Q_I\begin{bmatrix}\f_k\\ \j_j\end{bmatrix}
           =\begin{bmatrix}\Dt\j_j\\ i\f_k\end{bmatrix}$
  & \begin{picture}(5,9)(-1,5)
     \put(0,0){\includegraphics[height=11mm]{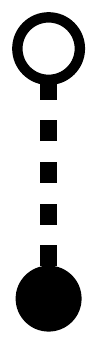}}
     \put(3,0){\scriptsize$j$}
     \put(3,9){\scriptsize$k$}
     \put(-1,4.5){\scriptsize$I$}
    \end{picture}\vrule depth4mm width0mm
     &  $Q_I\begin{bmatrix}\f_k\\ \j_j\end{bmatrix}
           =\begin{bmatrix}-\Dt\j_j\\-i\f_k\end{bmatrix}$ \\[5mm]
    \hline
  \MC4c{\footnotesize\baselineskip11pt
   The $I$-labeled edges may also be simply drawn in the $I^{\text{th}}$ color.}
  \end{tabular}
  \label{t:A}
\end{table}
The efficiency of this graphical encoding of the $Q$-action becomes obvious already in the following two examples of $N{=}2$ supermultiplets:
\begin{gather}
 \begin{aligned}
 \C3{Q_1}\,\vf_1&=\c_1,
 & \C1{Q_2}\,\vf_1&=\c_2,\\*
 \C3{Q_1}\,\vf_2&=\c_2,
 & \C1{Q_2}\,\vf_2&=-\c_1,\\*
 \C3{Q_1}\,\c_1&=i\Dt\vf_1,
 & \C1{Q_2}\,\c_1&=-i\Dt\vf_2,\\*
 \C3{Q_1}\,\c_2&=i\Dt\vf_2,
  \begin{picture}(36,0)(-2,0)
  \put(12,2){\includegraphics[width=18mm]{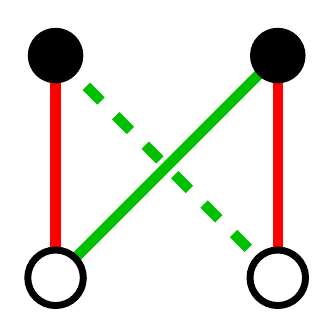}}
  \put(9,3){\small$\vf_1$}
  \put(9,18){\small$\c_1$}
  \put(29,3){\small$\vf_2$}
  \put(29,18){\small$\c_2$}
 \end{picture}\quad
 & \C1{Q_2}\,\c_2&=i\Dt\vf_1,
 \end{aligned}
 \label{e:N2b}\\[2mm]
 \begin{aligned}
 \C3{Q_1}\,\f&=\j_1,
 & \C1{Q_2}\,\f&=\j_2,\\*
 \C3{Q_1}\,\j_1&=i\Dt\f,
 & \C1{Q_2}\,\j_1&=-iF,\\*
 \C3{Q_1}\,\j_2&=iF,
 & \C1{Q_2}\,\j_2&=i\Dt\f,\\*
 \C3{Q_1}\,F&=\Dt\j_2,
  \begin{picture}(37,0)(-2,0)
  \put(13,-3){\includegraphics[width=18mm]{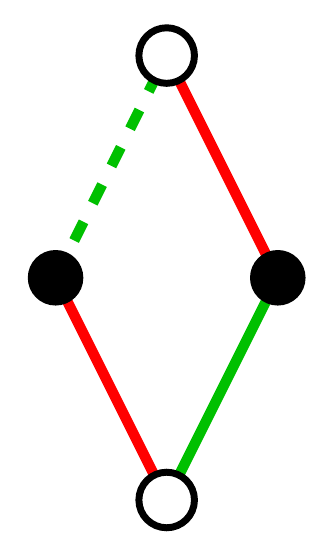}}
  \put(17,-1){\small$\f$}
  \put(10,11){\small$\j_1$}
  \put(30,11){\small$\j_2$}
  \put(24,22){\small$F$}
 \end{picture}\quad
 & \C1{Q_2}\,F&=-\Dt\j_1,
 \end{aligned}~ \label{e:N2t}
\end{gather}
The red (green) edges depict the action of $\C3{Q_1}$ ($\C1{Q_2}$), and it is fairly clear that a horizontal repositioning of the nodes is irrelevant, while lowering the highest node in\eq{e:N2t} produces a mapping from one to the other supermultiplet. Indeed, it is easy to see that the equations
\begin{equation}
  \f=\vf_1,\quad
  F=\Dt\vf_2,\quad
  \j_1=\c_1,\quad
  \j_2=\c_2,
\end{equation}
identify the supermultiplet\eq{e:N2b} with\eq{e:N2t}. This isomorphism is however not local since its inverse requires setting $\vf_2(\t)=\int_{\t_0}^\t\rd\t'\,F(\t')$, which depends on the values of $F(\t)$ globally over the full $[\t_0,\t]$-history of $F(\t)$---with a completely arbitrary lower limit, $\t_0$. The supermultiplets\eq{e:N2b} and\eq{e:N2t} must therefore be regarded as inequivalent---as their Adinkras indeed show.

Wherever possible, we thus use Adinkras to depict the corresponding $Q$-action within a supermultiplet rather than spell out all the equations, and will consequently also identify the supermultiplet with the Adinkra that depicts it.

Other classification results obtained in Refs.~\cite{r6-3c,r6-3,r6-3.2,r6-3.1} are less obvious: All Adinkras depicting supermultiplets of $N$-extended worldline supersymmetry must be $N$-dimensional hypercubes or iterated $\ZZ_2$-projections thereof, where all the $\ZZ_2$-projections are encoded by certain binary, error-detecting and error-correcting encryption codes. 
 Ref.~\cite{r6-3.2} reports that the number of possible supermultiplets grows combinatorially with $N$, and expects $\gtrsim10^{12}$ inequivalent Adinkras for $N\leq32$---and that's before taking into account the distinctions provided by the different height rearrangements of nodes.

Although these numbers are very large, they are finite. Furthermore, they imply that no Adinkra has more than $2^N$ nodes, for any fixed number of supercharges $N$, and so puts an upper limit on the size of supermultiplets that are depictable by Adinkras. Since no non-abelian Lie algebra has an upper limit on the size of its unitary finite-dimensional representations, one expects the classification of Adinkras to be only a first step in classifying off-shell supermultiplets.

\paragraph{The Puzzle:} Ref.~\cite{r6-1} noted that the supersymmetric gauge-quotient
\begin{equation}
  \Bigg(\vC{\includegraphics[width=10mm]{N2T.pdf}}
         \oplus
          \vC{\includegraphics[width=10mm]{N2T.pdf}}\Bigg)
  \Bigg/
   \raisebox{-7mm}{\includegraphics[width=10mm]{N2B.pdf}}
  ~~=~~
   \raisebox{9.5mm}{\rotatebox{180}{\reflectbox{%
    \includegraphics[width=10mm]{N2B.pdf}}}}
 \label{e:N2Q}
\end{equation}
where an Adinkra\eq{e:N2b} has been gauged away from a direct sum (general linear combination) of two Adinkras\eq{e:N2t} again has the structure of an Adinkra, the one depicted on the right-hand side of\eq{e:N2Q}; this follows unambiguously simply from counting the degrees of freedom per engineering dimension, \ie, nodes per height.

Ref.~\cite{r6-1} also noted that the same construction is not as easily resolved already for $N=3$. Counting of degrees of freedom now implies
 $3\,(1|3|3|1) - (3|4|1|0) = (0|5|8|3)$,
which does not resolve if the gauge-quotient is
\begin{equation}
 \left.
  \left(\vC{\includegraphics[width=10mm]{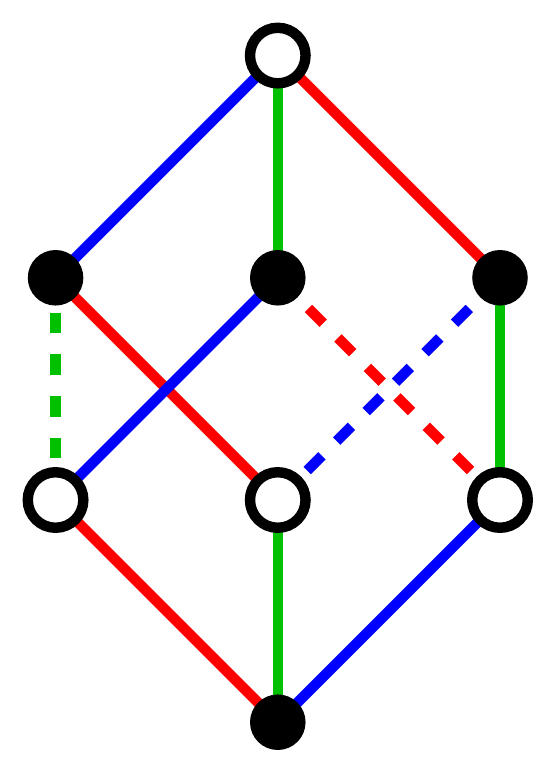}}
         \oplus
          \vC{\includegraphics[width=10mm]{N3T.pdf}}
           \oplus
            \vC{\includegraphics[width=10mm]{N3T.pdf}}\right)
  \right/\mkern-9mu
   \raisebox{-8mm}{\includegraphics[width=15mm]{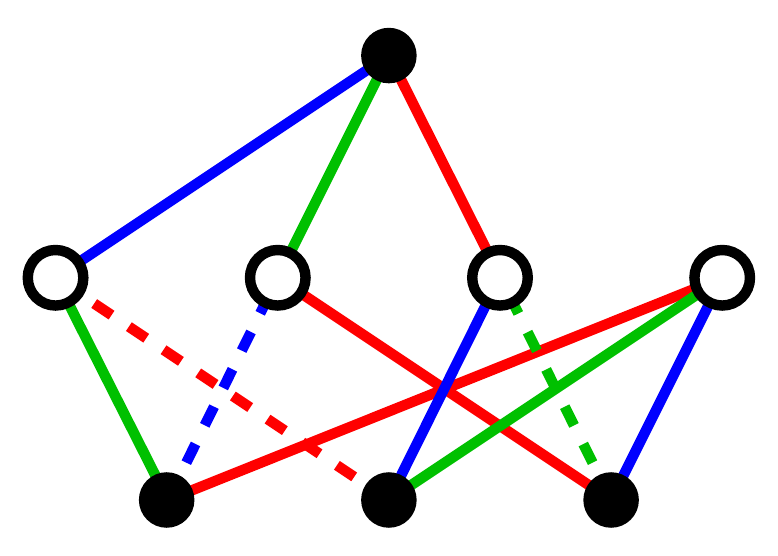}}
  \quad\overset{?}{=}~
   \raisebox{-2mm}{\includegraphics[width=13mm]{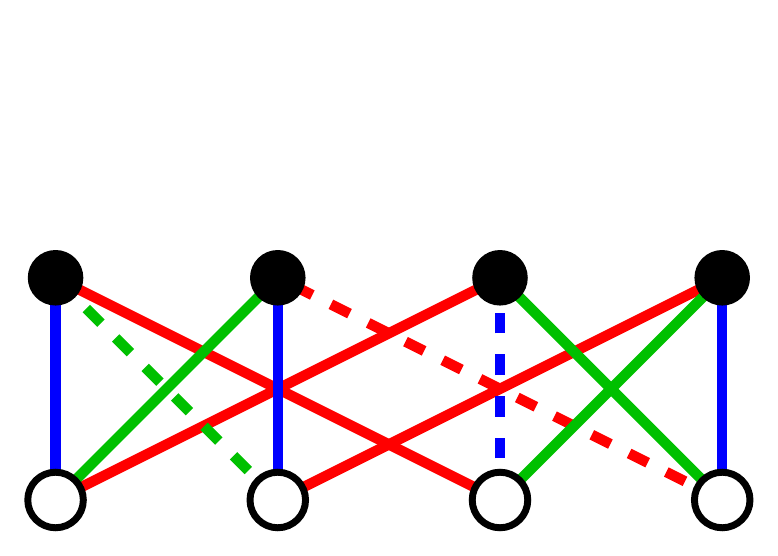}}
  \oplus
   \raisebox{-2mm}{\includegraphics[width=13mm]{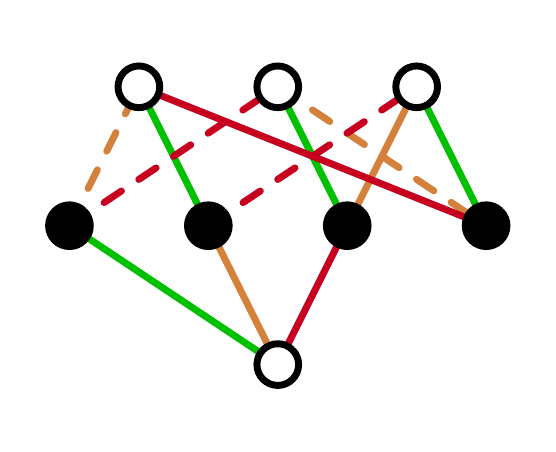}}
  ~\text{,\quad or}~\overset{?}{=}~
   \raisebox{-2mm}{\includegraphics[width=13mm]{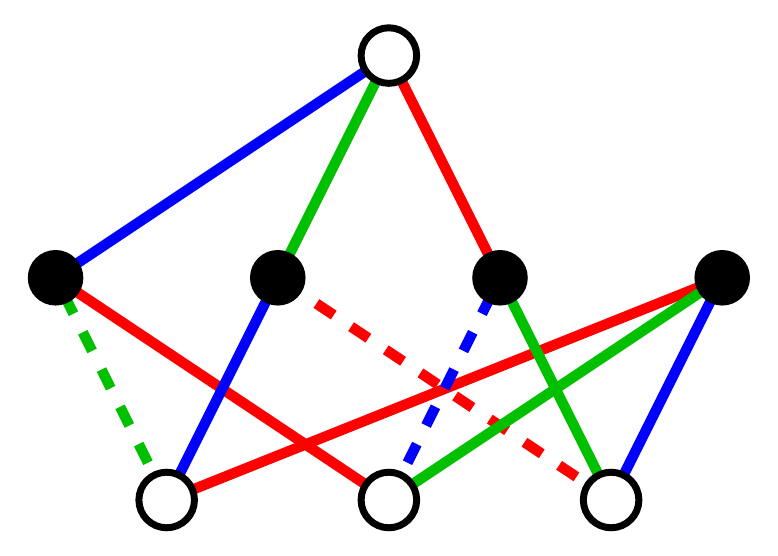}}
  \oplus
   \raisebox{-2mm}{\includegraphics[width=13mm]{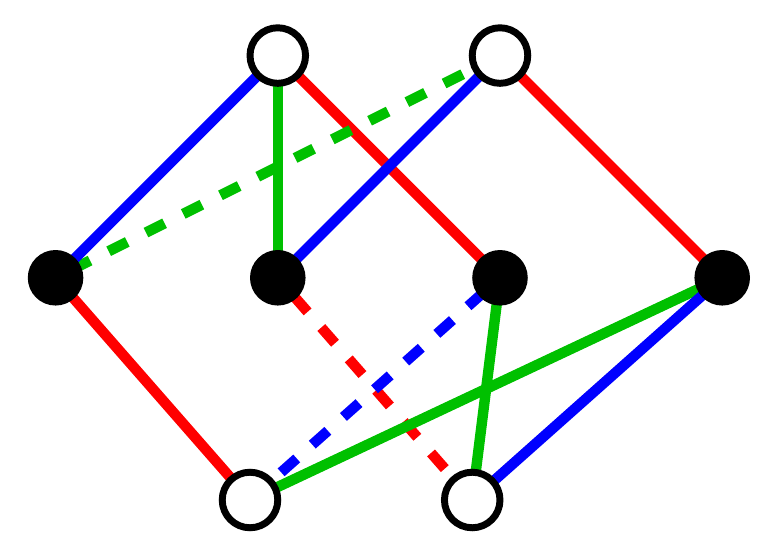}}
 \label{e:N3Q}
\end{equation}
Gauge-quotients are most familiar in physics from electromagnetism, where the 4-vector gauge potential is defined only up to the gauge transformation, $A_\mu \simeq A_\mu + \partial_\mu\lambda$. The Fourier transform of this indicates unambiguously that the component of $A_\mu$ along the 4-momentum of the photon, $k_\mu$, is an unphysical degree of freedom.

 It is then more than a little peculiar that in the analogous gauge-quotient\eq{e:N3Q}, the result is not as evidently and as unambiguously determined, but this question has been left open in Ref.~\cite{r6-1}, and has remained unanswered since then. Notably, the quotient\eq{e:N3Q} is also but the first non-trivial step in an indefinite sequence of constructions; see the display~(8.28) in Ref.\cite{r6-1} and the displays\eq{e:ZigZag} and\eq{e:ZigZagA}, below.

\section{The Structure of the Gauge-Quotient Supermultiplet}
\label{s:GQS}
To flesh out the details of the construction\eq{e:N3Q}, we need to introduce the two supermultiplets used therein (see appendix~\ref{s:X+DX}), and we fix the number of supersymmetries at $N=3$.

\begin{figure}[ht]
 \centering
  \begin{picture}(150,40)(0,2)
   \put(.5,0){\includegraphics[width=152mm]{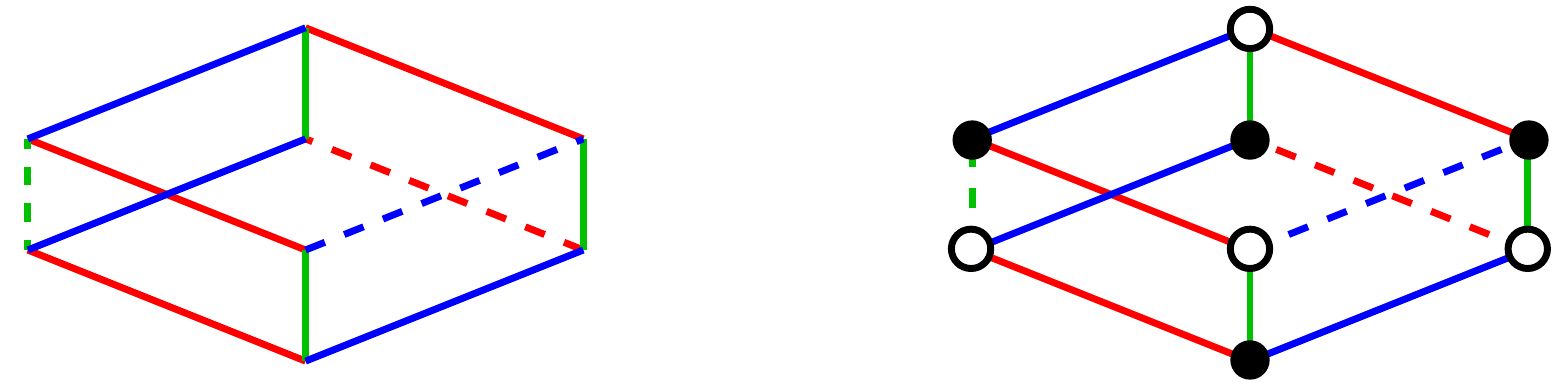}}
     \put(30,35){\cB{$\cY_k$}}
     \put(3,24){\bB{$\Y_k{}^3$}}
     \put(30,24){\bB{$\Y_k{}^2$}}
     \put(56,24){\bB{$\Y_k{}^1$}}
     \put(3,13){\cB{$Y_{k1}$}}
     \put(30,13){\cB{$Y_{k2}$}}
     \put(56,13){\cB{$Y_{k3}$}}
     \put(30,2){\bB{$\eta_k$}}
     \put(68,20){or simply}
 \end{picture}
 \caption{The $k^\text{th}$ fermionic intact supermultiplet, represented also by the superfield $\IY_k$}
 \label{f:Yk}
\end{figure}

\begin{figure}[ht]
 \centering
  \begin{picture}(150,42)(2,0)
   \put(.5,0){\includegraphics[width=160mm]{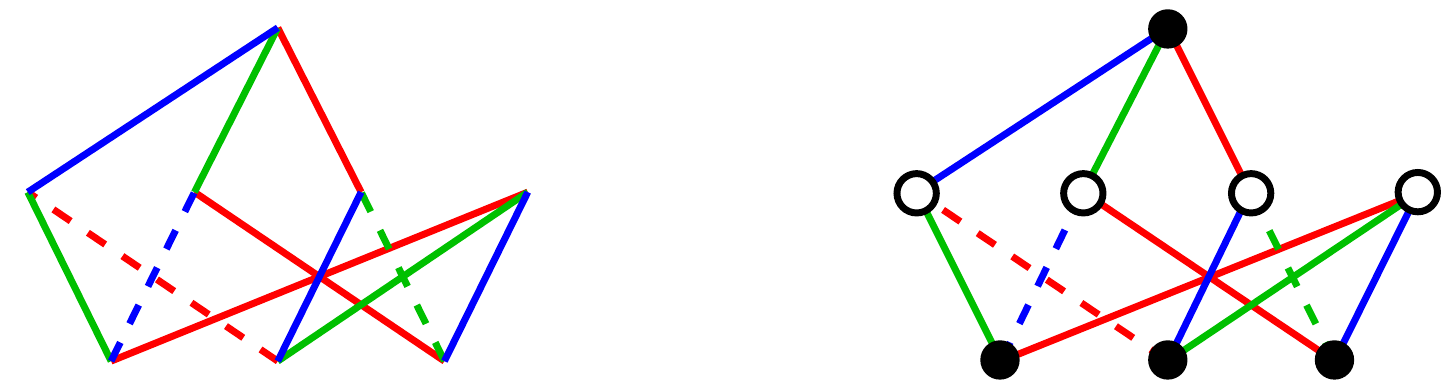}}
     \put(31,38){\bB{$\L$}}
     \put(3,21){\cB{$L^3$}}
     \put(22,21){\cB{$L^2$}}
     \put(40,21){\cB{$L^1$}}
     \put(58,21){\cB{$L^0$}}
     \put(13,2){\bB{$\lambda_1$}}
     \put(31,2){\bB{$\lambda_2$}}
     \put(50,2){\bB{$\lambda_3$}}
     \put(72,20){or simply}
 \end{picture}
 \caption{A supermultiplet represented by the superfield
 $\IL_I\protect\Defl i\rD_I\IX$, for $\IX$ a bosonic intact superfield}
 \label{f:DX}
\end{figure}

\paragraph{The ``Ingredients'':}
We will need a triple of {\em\/intact\/}\footnote{Herein, ``intact'' is short for ``unconstrained, ungauged, unprojected and otherwise unrestricted.''} fermionic supermultiplets, shown in figure~\ref{f:Yk},
where $k=1,2,3$ counts the three copies at the left-hand side of\eq{e:N3Q}, and from which the $Q$-action
\begin{equation}
 \IY_k:\qquad
  \begin{array}{@{} c@{:~~}c|ccc|ccc|c @{}}
 \omit   & \eta_k    & Y_{k1}    & Y_{k2}    & Y_{k3}
          & \Y_k{}^1 & \Y_k{}^2 & \Y_k{}^3 & \cY_k \\[1pt]
    \hline\rule{0pt}{2.5ex}
\C3{Q_1}
 & iY_{k1} & \Dt\eta_k   & \Y_k{}^3  & -\Y_k{}^2
          & i\cY_k       & -i\Dt{Y}_{k3}  & i\Dt{Y}_{k2} & \Dt\Y_k{}^1 \\ 
\C1{Q_2}
 & iY_{k2} & -\Y_k{}^3 & \Dt\eta_k   & \Y_k{}^1
          & i\Dt{Y}_{k3}  & i\cY_k        & -i\Dt{Y}_{k1}  & \Dt\Y_k{}^2 \\ 
\color{blue}{Q_3}
 & iY_{k3} & \Y_k{}^2  & -\Y_k{}^1 & \Dt\eta_k
          & -i\Dt{Y}_{k2} & i\Dt{Y}_{k1}   & i\cY_k   & \Dt\Y_k{}^3 \\ 
    \hline
  \end{array}
 \label{e:QTY}
\end{equation}
may be read of. This supermultiplet coincides with the supermultiplet encoded by the standard fermionic {\em\/superfield\/}, $\IY_k$, as originally introduced by Salam and Strathdee~\cite{rSSSS4}, and is routinely used in supersymmetry texts~\cite{r1001,rPW,rWB,rBK}.
 We will also need a second supermultiplet, depicted in figure~\ref{f:DX}, 
graphically encoding the transformation rules
\begin{equation}
 \IL_I\id i\rD_I\IX:\qquad
  \begin{array}{@{} c@{:~~}ccc|cccc|c @{}}
 \omit& \lambda_1 & \lambda_2 & \lambda_3 & L^0 & L^1 & L^2 & L^3 & \L \\[1pt] 
    \hline\rule{0pt}{2.5ex}
\C3{Q_1}
 &  iL^0 & -iL^3 &  iL^2 & \Dt\lambda_1 &  \L      &  \Dt\lambda_3 & -\Dt\lambda_2 & i\Dt{L}^1 \\ 
\C1{Q_2}
 &  iL^3 &  iL^0 & -iL^1 & \Dt\lambda_2 & -\Dt\lambda_3 &  \L      &  \Dt\lambda_1 & i\Dt{L}^2 \\ 
\color{blue}{Q_3}
 & -iL^2 &  iL^1 &  iL^0 & \Dt\lambda_3 &  \Dt\lambda_2 & -\Dt\lambda_1 &  \L      & i\Dt{L}^3 \\ 
    \hline
  \end{array}
 \label{e:QTDX}
\end{equation}
This supermultiplet is in fact a {\em\/superderivative\/} $(i\rD_I\IX)$ of a standard, intact bosonic superfield, $\IX$.

\paragraph{The Gauge-Quotient:}
Armed with the transformations\eq{e:QTY} and\eq{e:QTDX} and the corresponding superfield notation and calculus~\cite{r1001,rPW,rWB,rBK}, we are now in the position to compute
\begin{equation}
  \cok\big(\IX\7{i\rD_I}{\into}\IY_I\big) =
  \big(\IY_I\big/i\rD_I\IX\big),
   \quad\ie,\quad
  \IY_I\simeq\IY_I+\IL_I,
   \quad\text{or}\quad
  \delta\IY_I=\IL_I\Defl i\rD_I\IX.
 \label{e:Y/iDX}
\end{equation}
where we have identified the values of $k=1,2,3$ with the values of $I=1,2,3$. The component fields (4 bosons + 4 fermions) of the supermultiplet $\IL_I\Defl(i\rD_I\IX)$ are thus identified as the gauge parameters that can be used to eliminate some of the (linear combinations of the) $12+12$ component fields in $\IY_I$. Notice the similarity between\eq{e:Y/iDX} and $A_\m\simeq A_\m+\vd_\m\l$ in electromagnetism.
 The component-level content of the gauge transformation $\delta\IY_I= \IL_I$ is shown in table~\ref{t:dY=L}, along with the superderivatives used in the component projections.
\begin{table}[ht]
 \caption{\small\baselineskip=12pt The components of the gauge transformation $\d\IY_I=\IL_I$}
 \vspace{4mm}
\centerline{$
  \begin{array}{@{} r@{:~}r@{\>=\>}lr@{\>=\>}lr@{\>=\>}l @{}}
 \text{Proj.}& \d\IY_1 & \IL_1 & \d\IY_2 & \IL_2 & \d\IY_3 & \IL_3 \\[1pt]
    \hline\rule{0pt}{2.9ex}
 \Ione & \d\h_1 & \l_1 & \d\h_2 & \l_2 & \d\h_3 & \l_3 \\[1pt]
    \hline\rule{0pt}{2.9ex}
 -\rD_1 & \d Y_{11} &  L^0 & \d Y_{21} & -L^3 & \d Y_{31} &  L^2 \\[1pt]
 -\rD_2 & \d Y_{12} &  L^3 & \d Y_{22} &  L^0 & \d Y_{32} & -L^1 \\[1pt]
 -\rD_3 & \d Y_{13} & -L^2 & \d Y_{23} &  L^1 & \d Y_{33} &  L^0 \\
    \hline
  \end{array}
  \quad
  \begin{array}{@{} r@{:~}r@{\>=\>}lr@{\>=\>}lr@{\>=\>}l @{}}
 \text{Proj.}& \d\IY_1 & \IL_1 & \d\IY_2 & \IL_2 & \d\IY_3 & \IL_3 \\[1pt]
    \hline\rule{0pt}{2.9ex}
 -i\rD_1\rD_2 & \d\Y_1{}^3 & -\Dt\l_2   & \d\Y_2{}^3
              &  \Dt\l_1   & \d\Y_3{}^3 & -\L \\
  i\rD_1\rD_3 & \d\Y_1{}^2 &  \Dt\l_3   & \d\Y_2{}^2
              & -\L        & \d\Y_3{}^2 & -\Dt\l_1 \\
 -i\rD_2\rD_3 & \d\Y_1{}^1 & -\L        & \d\Y_2{}^1
              & -\Dt\l_3   & \d\Y_3{}^1 &  \Dt\l_2 \\
    \hline\rule{0pt}{2.9ex}
 i\rD_1\rD_2\rD_3 & \d\cY_1 & -\Dt{L}{}^1 & \d\cY_2 & -\Dt{L}{}^2 & \d\cY_3 & -\Dt{L}{}^3 \\
    \hline
  \end{array}
$}
 \label{t:dY=L}
\end{table}

The entries in the $\Ione$-row offer no freedom of choice in eliminations\ft{The gauge-quotienting identifies fermions in a way that reminds of the similar identification used in Ref.\cite{rBKLS-SFN4}. However, that analysis {\em\/begins\/} with fermion identifications, whereas component field identifications here emerges as a consequence of gauge-quotienting. This connects the two approaches and may well permit an extension of the present analysis also to nonlinear representations of supersymmetry; we thank the Referee for pointing this out.}:
\begin{equation}
  \l_I = -\h_I, \qquad\text{whereby}\qquad \h_I\simeq 0.
 \label{e:L=-Y0}
\end{equation}
This eliminates the lowest component fermions from the $\IY_I/\IL_I$ gauge equivalence class. 
 Applying this gauge also fixes all the ``off-diagonal'' higher-level fermions to appear only in the binomials:
\begin{subequations}
 \label{e:6odYs}
\begin{alignat}9
  \Y_1{}^2&\mapsto(\Y_1{}^2{-}\Dt\h_3)&&\Defr\bar\Y_1{}^2,&\qquad
  \Y_1{}^3&\mapsto(\Y_1{}^3{+}\Dt\h_2)&&\Defr\bar\Y_1{}^3,&\qquad
  \Y_2{}^3&\mapsto(\Y_2{}^3{-}\Dt\h_1)&&\Defr\bar\Y_2{}^3,\\
  \Y_2{}^1&\mapsto(\Y_2{}^1{+}\Dt\h_3)&&\Defr\bar\Y_2{}^1,&\qquad
  \Y_3{}^1&\mapsto(\Y_3{}^1{-}\Dt\h_2)&&\Defr\bar\Y_3{}^1,&\qquad
  \Y_3{}^2&\mapsto(\Y_3{}^2{+}\Dt\h_1)&&\Defr\bar\Y_3{}^2.
\end{alignat}
\end{subequations}
This (intermediate) field redefinition reduces the components of $(\IY_I/\IL_I)$ to the nine fermions
\begin{equation}
  \begin{bmatrix}
        (\Y_1{}^1{-}\L)  & \bar\Y_1{}^2     & \bar\Y_1{}^3 \\
        \bar\Y_2{}^1     & (\Y_2{}^2{-}\L)  & \bar\Y_2{}^3 \\
        \bar\Y_3{}^1     & \bar\Y_3{}^2      &(\Y_3{}^3{-}\L)
  \end{bmatrix}
\end{equation}
and twelve bosons:
\begin{equation}
 \left(
  \begin{bmatrix}
        (Y_{11}{+}L^0) & (Y_{21}{-}L^3) & (Y_{31}{+}L^2) \\[3pt]
       (Y_{12}{+}L^3) & (Y_{22}{+}L^0) & (Y_{32}{-}L^1) \\[3pt]
       (Y_{13}{-}L^2) & (Y_{23}{+}L^1) & (Y_{33}{+}L^0)
  \end{bmatrix}~\right|\left.
  \begin{bmatrix}
        (\cY_1{-}\Dt{L}{}^1) \\ (\cY_2{-}\Dt{L}{}^2) \\ (\cY_3{-}\Dt{L}{}^3)
  \end{bmatrix}\right).
\end{equation}
We remain with a free choice of $\L$, as well as $L^0,L^1,L^2,L^3$. These gauge parameters may be used to eliminate one linear combination each:
\begin{equation}
   \L\mapsto\sum_I \ell_I{}^I\Y_I{}^I,
 \qquad
   L^0\mapsto\sum_I \ell^{II} Y_{II},
 \qquad
   L^I\mapsto\sum_{\sss I\neq J\neq K\neq I}\ell^{IJK} Y_{JK}.
 \label{e:72}
\end{equation}
Among this {\em\/continuum\/} of gauge-choices, we focus on those that are most likely to simplify the supersymmetry transformation rules within the resulting gauge-quotient representative.

\paragraph{A Gauge-Choice:}
One such gauge-choice,
\begin{equation}
  \L=\Y_3{}^3,~~
  L^0=-Y_{33},~~
  L^1=+Y_{32},~~
  L^2=-Y_{31}~~\text{and}~~
  L^3=+Y_{21}
 \label{e:L=-Y1}
\end{equation}
and the particular accompanying field redefinitions\footnote{Field redefinitions are restricted by the engineering dimensions\eq{e:EDY} and locality.}
\begin{subequations}
\label{e:SimY}
\begin{alignat}9
  \eU_1{}^1&\Defl\Y_1{}^1{-}\Y_3{}^3,&\quad
  \eU_2{}^2&\Defl\Y_2{}^2{-}\Y_3{}^3,&\quad
  \eU_1{}^3&\Defl\Y_1{}^3{+}\Y_3{}^1,&\quad
  \eU_2{}^3&\Defl\Y_2{}^3{+}\Y_3{}^2;\\*
  \eU_1{}^2&\Defl\Y_1{}^2{-}\Dt\h_3,&\quad
  \eU_2{}^1&\Defl\Y_2{}^1{+}\Dt\h_3,&\quad
  \eU_3{}^1&\Defl\Y_3{}^1{-}\Dt\h_2,&\quad
  \eU_3{}^2&\Defl\Y_3{}^2{+}\Dt\h_1;\\
 \eY_{11}&\Defl Y_{11}{-}Y_{33},&\quad
 \eY_{12}&\Defl Y_{12}{+}Y_{21},&\quad
 \eY_{13}&\Defl Y_{13}{+}Y_{31},&\quad
 \eY_{23}&\Defl Y_{23}{+}Y_{32},\\*
 \eY_{22}&\Defl Y_{22}{-}Y_{33},&\quad
 \EY_1&\Defl \cY_1{-}\Dt{Y}_{32},&\quad
 \EY_2&\Defl \cY_2{+}\Dt{Y}_{31},&\quad
 \EY_3&\Defl \cY_3{-}\Dt{Y}_{21}
\end{alignat}
\end{subequations}
jointly lead to 
\begin{equation}
  \begin{array}{@{} c|ccc @{}}
 & \C3{Q_1} & \C1{Q_2} & \color{blue}{Q_3} \\ 
    \hline\rule{0pt}{2.9ex}
\eU_1{}^1
 & i\EY_1 & i\Dt\eY_{13} & -i\Dt\eY_{12}\C5{{-}i\EY_3} \\[0pt]
\eU_2{}^2
 & -i\Dt\eY_{23} & i\EY_2 & -i\EY_3 \\[0pt]
\eU_1{}^2
 & -i\Dt\eY_{13} & i\EY_1 & i\Dt\eY_{11} \\[0pt]
\eU_2{}^1
 & i\EY_2 & i\Dt\eY_{23} & -i\Dt\eY_{22} \\[0pt]
\eU_1{}^3
 & i\Dt\eY_{12}\C5{{+}i\EY_3} & -i\Dt\eY_{11} & i\EY_1 \\[0pt]
\eU_3{}^1
 & i\EY_3 & -i\Dt\eY_{22} & -i\Dt\eY_{23} \\[0pt]
\eU_2{}^3
 & i\Dt\eY_{22} & i\EY_3 & i\EY_2 \\[0pt]
\eU_3{}^2
 & i\Dt\eY_{11} & i\Dt\eY_{12}\C5{{+}i\EY_3} & i\Dt\eY_{13} \\ 
    \hline
  \end{array}
 \quad
  \begin{array}{@{} c|ccc @{}}
 & \C3{Q_1} & \C1{Q_2} & \color{blue}{Q_3} \\ 
    \hline\rule{0pt}{2.1ex}
\eY_{11}
 & \eU_3{}^2 & -\eU_1{}^3 & \eU_1{}^2 \\[0pt]
\eY_{22}
 & \eU_2{}^3 & -\eU_3{}^1 & -\eU_2{}^1 \\[0pt]
\eY_{12}
 & \eU_1{}^3\C5{{-}\eU_3{}^1} & \eU_3{}^2\C5{{-}\eU_2{}^3}
   & -\eU_1{}^1\C5{{+}\eU_2{}^2} \\[-2pt]
\eY_{13}
 & -\eU_1{}^2 & \eU_1{}^1 & \eU_3{}^2 \\[0pt]
\eY_{23}
 & -\eU_2{}^2 & \eU_2{}^1 & -\eU_3{}^1 \\
    \cline{2-4}\rule{0pt}{3.1ex}
\EY_1
 & \Dt\eU_1{}^1 & \Dt\eU_1{}^2 & \Dt\eU_1{}^3 \\[0pt]
\EY_2
 & \Dt\eU_2{}^1 & \Dt\eU_2{}^2 & \Dt\eU_2{}^3 \\[0pt]
\EY_3
 & \Dt\eU_3{}^1 & \Dt\eU_2{}^3 & -\Dt\eU_2{}^2 \\
    \hline
  \end{array}
  \label{e:QY-L3}
\end{equation}
which permits no further simplification: row-operations in the left-hand half of the table\eq{e:QY-L3} imply field substitutions in the right-hand half. For example, adding the $\eU_3{}^1$-row to the $\eU_1{}^3$-row so as to cancel out the $i\EY_3$ entry in $\C3{Q_1}(\eU_1{}^3)$ would force a field redefinition
\begin{equation}
  \eU_1{}^3\mapsto\Tw{\Y}_1{}^3\Defl(\eU_1{}^3{-}\eU_3{}^1),
   \quad\text{and so then}\quad
  \eU_1{}^3\mapsto\Tw{\Y}_1{}^3+\eU_3{}^1,
\end{equation}
which then must be substituted throughout the right-hand half of table~\eq{e:QY-L3}. While this eliminates the one binomial entry in $\C3{Q_1}(\eU_1{}^3)$ as well as the one in $\C3{Q_1}(\eY_{12})$, it creates binomial entries for $\C1{Q_2}(\eU_1{}^3)$ and $\C6{Q_3}(\eU_1{}^3)$, as well as in $\C1{Q_2}(\eY_{11})$ and $\C6{Q_3}(\EY_1)$. The net effect then is a basis of component fields for which system of $Q$-transformations in $\IY_I/(i\rD_I\IX)$ is {\em\/more\/} rather than less complicated. In this, practical sense is the component field basis
\begin{equation}
  \IY_I/(i\rD_I\IX):~~
   \big\{\, \eY_{11},\eY_{22},\eY_{12},\eY_{13},\eY_{23} \,\big|
          \, \eU_1{}^1,\eU_2{}^2,\eU_1{}^2,\eU_2{}^1,
              \eU_1{}^3,\eU_3{}^1,\eU_2{}^3,\eU_3{}^2 \,\big|
            \, \EY_1,\EY_2,\EY_3 \,\big\}
 \label{e:Y/iDX2}
\end{equation}
an optimal choice.

\begin{figure}[ht]
\centering
 \begin{picture}(140,45)
   \put(0,0){\includegraphics[width=140mm]{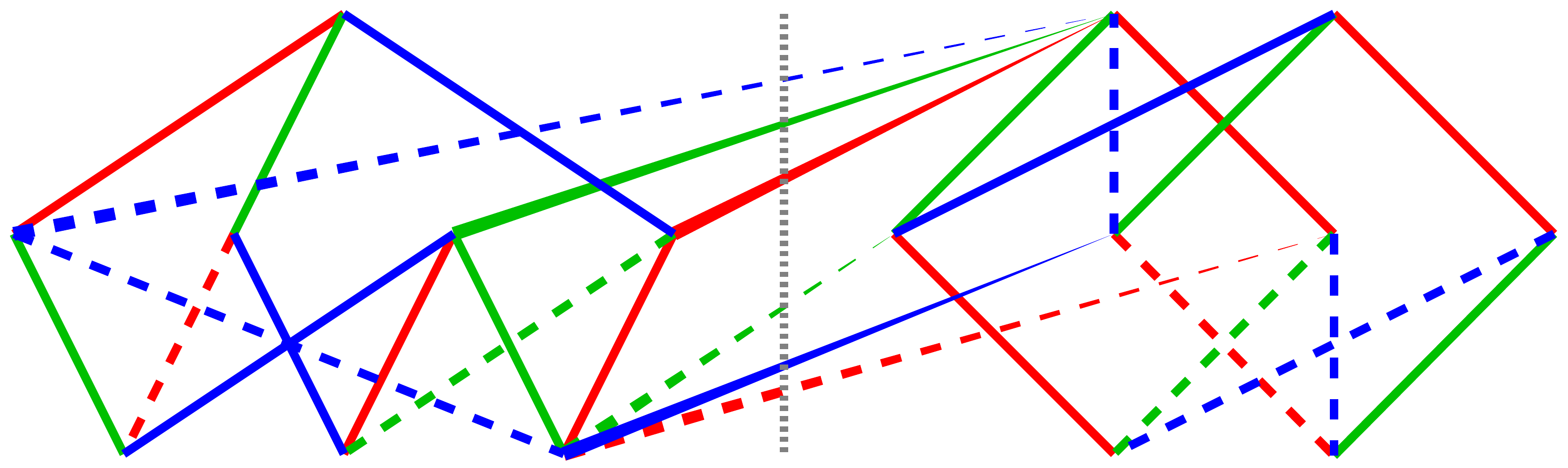}}
    \put(12,1){\cB{$\eY_{13}$}}
    \put(31.5,1){\cB{$\eY_{11}$}}
    \put(51,1){\cB{$\eY_{12}$}}
    \put(99.5,1){\cB{$\eY_{22}$}}
    \put(119,1){\cB{$\eY_{23}$}}
    \put(2,20){\bB{$\eU_1{}^1$}}
    \put(21,20){\bB{$\eU_1{}^2$}}
    \put(40,20){\bB{$\eU_3{}^2$}}
    \put(59,20){\bB{$\eU_1{}^3$}}
    \put(80,20){\bB{$\eU_2{}^3$}}
    \put(99,20){\bB{$\eU_2{}^2$}}
    \put(118,20){\bB{$\eU_3{}^1$}}
    \put(137,20){\bB{$\eU_2{}^1$}}
    \put(31,40){\cB{$\EY_1$}}
    \put(99.5,40){\cB{$\EY_3$}}
    \put(119,40){\cB{$\EY_2$}}
 \end{picture}
\caption{An Adinkra-like depiction of the gauge-quotient supermultiplet $\IY_I/(i\rD_I\IX)$}
 \label{f:A>B}
\end{figure}

 The result\eq{e:QY-L3} is depicted in figure~\ref{f:A>B}, akin to an Adinkra.
This depiction contains a novel graphic element as compared to the by now standard Adinkra elements in table~\ref{t:A}: tapering edges. These reflect the ``one-way'' nature of the depicted $Q$-action, where for example $\C3{Q_1}(\eY_{12})\supset\eU_3{}^1$, but $\C3{Q_1}(\eU_3{}^1)\not\supset\Dt\eY_{12}$. This type of ``one-way'' (partial) $Q$-action, shown in\eq{e:QY-L3} using pale blue ink, was noted in unrelated studies only in rather more complicated supermultiplets~\cite{r6-4.2,rGHHS-CLS}.

\paragraph{Other Gauge-Choices:}
The continuous palette of gauge-choices\eq{e:72} parametrizes a continuous family of component field bases, and the optimizing experimentations\eqs{e:L=-Y1}{e:Y/iDX2} leading to the graph in figure~\ref{f:A>B} clearly reaffirm that some basis choices are much more optimal than others.

Fortunately, a similarly detailed exploration of the continuously many remaining gauge-choices\eq{e:72} turns out to be unnecessary in addressing the two main questions:
\begin{enumerate}\itemsep=-3pt\vspace{-2mm}
 \item Is there a gauge-choice that splits $\IY_I/(i\rD_I\IX)$ into a direct sum of two Adinkras\eq{e:N3Q}?
 \item Is there a gauge-choice wherein
 $\IY_I/(i\rD_I\IX)=(4|4|0)\underrightarrow{\oplus}(1|4|3)$, \ie, wherein $\IY_I/(i\rD_I\IX)$ exhibits the structure of a $(4|4|0)$- and a $(1|4|3)$-component Adinkra connected by one-way $Q$-action indicated by the under-arrow in ``$\underrightarrow{\oplus}$''?
\end{enumerate}
To this end, it actually suffices to explore $Q_I(\cY_J)$. For example,
\begin{equation}
 \vC{\begin{picture}(140,25)(0,-2)
 \setlength\fboxsep{1pt}
   \put(0,0){\includegraphics[width=140mm]{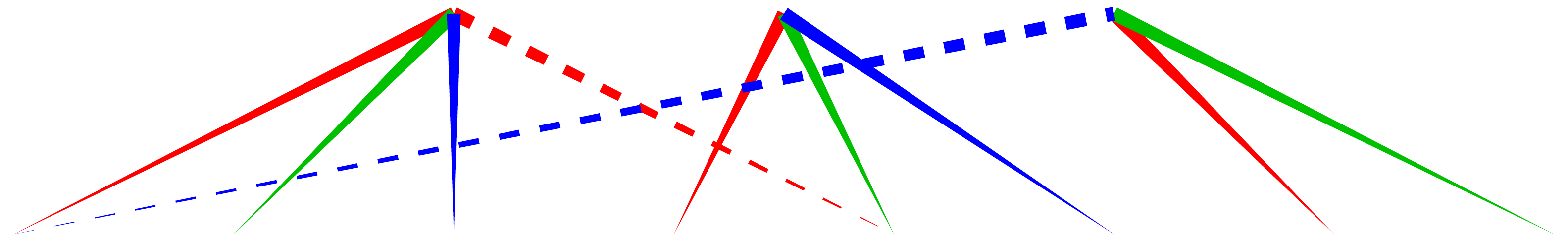}}
    \put(2,0){\bB{${\tw\Y}_1{}^1$}}
    \put(21,0){\bB{${\tw\Y}_1{}^2$}}
    \put(40,0){\bB{${\tw\Y}_1{}^3$}}
    \put(59,0){\bB{${\tw\Y}_2{}^1$}}
    \put(80,0){\bB{${\tw\Y}_2{}^2$}}
    \put(99,0){\bB{${\tw\Y}_2{}^3$}}
    \put(118,0){\bB{${\tw\Y}_3{}^1$}}
    \put(137,0){\bB{${\tw\Y}_3{}^2$}}
    \put(40,19){\cB{$\tw\cY_1$}}
    \put(71,19){\cB{$\tw\cY_3$}}
    \put(100,19){\cB{$\tw\cY_2$}}
 \end{picture}}
 \label{e:A^B}
\end{equation}
is obtained by employing the gauge-choice $\L=\Y_3{}^3$, $(L^0,L^1,L^2,L^3)=(-Y_{22},-Y_{23},-Y_{31},-Y_{12})$, where $\tw\Y_k{}^I$ and $\tw\cY_k$ redefine these component fields so as to minimize the number of edges in\eq{e:A^B}.
 The edges are all drawn tapering to indicate that the ``reverse'' supersymmetry transformations from $Q_I(\tw\Y_J{}^K)$ are not shown---nor do they need to be computed.
 The partial graph\eq{e:A^B} already shows that $\IY_I/(i\rD_I\IX)$ in this gauge-choice cannot separate so as to permit a direct sum decomposition for question~1.
 Answering question~2 in turn, four of the fermions should have been omitted in the results of $Q_I(\tw\cY_J)$ for a $(4|4|0)\underrightarrow{\oplus}(1|4|3)$ structure, but not one is so omitted.

There remains the logical possibility of gauge-choices wherein the otherwise proper $(4|4|0)$- and $(1|4|3)$-node Adinkras are connected only by one-way transformations. This could include cases where the one-way $Q$-action is directed only downward, or maybe in both directions, just not in the $\cY_I\to\Y_I{}^J$ direction checked by\eq{e:A^B}. To eliminate these as well as the cases in which the otherwise proper $(3|4|1)$- and $(2|4|2)$-node Adinkras are connected differently than in\eq{f:A>B}, symbolically denoted
\begin{equation}
 (4|4|0)\overleftarrow{\oplus}(1|4|3),\quad
 (4|4|0)\overleftarrow{\underrightarrow{\oplus}}(1|4|3),\quad
 (3|4|1)\overleftarrow{\oplus}(2|4|2),\quad
 (3|4|1)\overleftarrow{\underrightarrow{\oplus}}(2|4|2),
\end{equation}
one must additionally also compute the $Q_I(Y_{JK})$ transformations and perhaps a few of the $Q_I(\Y_J{}^K)$ transformations. This analysis is still simpler than computing the transformations depending on the numerous and continuous $\ell$-coefficients from\eq{e:72}, and reaffirms the conclusion that the gauge-quotient $\IY_I/(i\rD_I\IX)$ does not decompose, and does include one-way $Q$-action in every possible gauge-choice.

\section{An Indefinite Sequence of Representations}
\label{s:ZZ}
All by itself, the particular example\eqs{e:Y/iDX}{e:Y/iDX2} may be thought of as fairly unremarkable, and especially so from the point of view of any possible immediate physics application:
 ({\small\bf1})~The number ($N\,{=}\,3$) of worldline supersymmetries does not relate to supersymmetry in higher-dimensional spacetimes other than some worldsheet models\cite{rGH-obs,rH-WWS}, and in particular cannot be extended to the most interesting physics applications, in $1{+}3$-dimensional (or larger) spacetime.
 ({\small\bf2})~The gauge-equivalence class $\IY_I/(i\rD_I\IX)$ itself does not, to the best of our knowledge, appear in any of the known physics models.

Nevertheless, the particular example\eqs{e:Y/iDX}{e:Y/iDX2} turns out to be the simplest in an indefinite sequence of such ever larger gauge-equivalence quotients, and exhibits the properties discussed above, which we now argue are completely generic for $N\geq3$.

To this end, we must recall a completely general construction proposed in Ref.\cite{r6-1}, and aided by a concept precisely defined in appendix~B of Ref.\cite{r6-3.2}:
\begin{defn}\label{linalg}\addtolength{\rightskip}{2pc}
A {\bsf\/strict homomorphism\/} of off-shell supermultiplets is a supersymmetry-preserving linear map $\m: \IM_1 \to \IM_2$, such that the quotient $\IM_2 / \m(\IM_1)=\cok(\m)$ is also an off-shell supermultiplet.
\end{defn}
The gauge-equivalence\eq{e:Y/iDX} indeed defines an off-shell supermultiplet, and has been obtained as the quotient of the supermultiplet $\IY_I$ given in\eq{e:QTY} and depicted in figure~\ref{f:Yk},  by (the imbedding of) the supermultiplet $\IL_I\Defl(i\rD_I\IX)$ given in\eq{e:QTDX} and depicted in figure~\ref{f:DX}. In this precise sense, we identify
\begin{equation}
 \left.
  \left(\vC{\includegraphics[height=20mm]{N3T.pdf}}
         \oplus
          \vC{\includegraphics[height=20mm]{N3T.pdf}}
           \oplus
            \vC{\includegraphics[height=20mm]{N3T.pdf}}\right)
  \right/\mkern-9mu
   \raisebox{-8.5mm}{\includegraphics[height=14mm]{N3F341.pdf}}
  \quad=\quad
   \raisebox{-3mm}{\includegraphics[height=14mm]{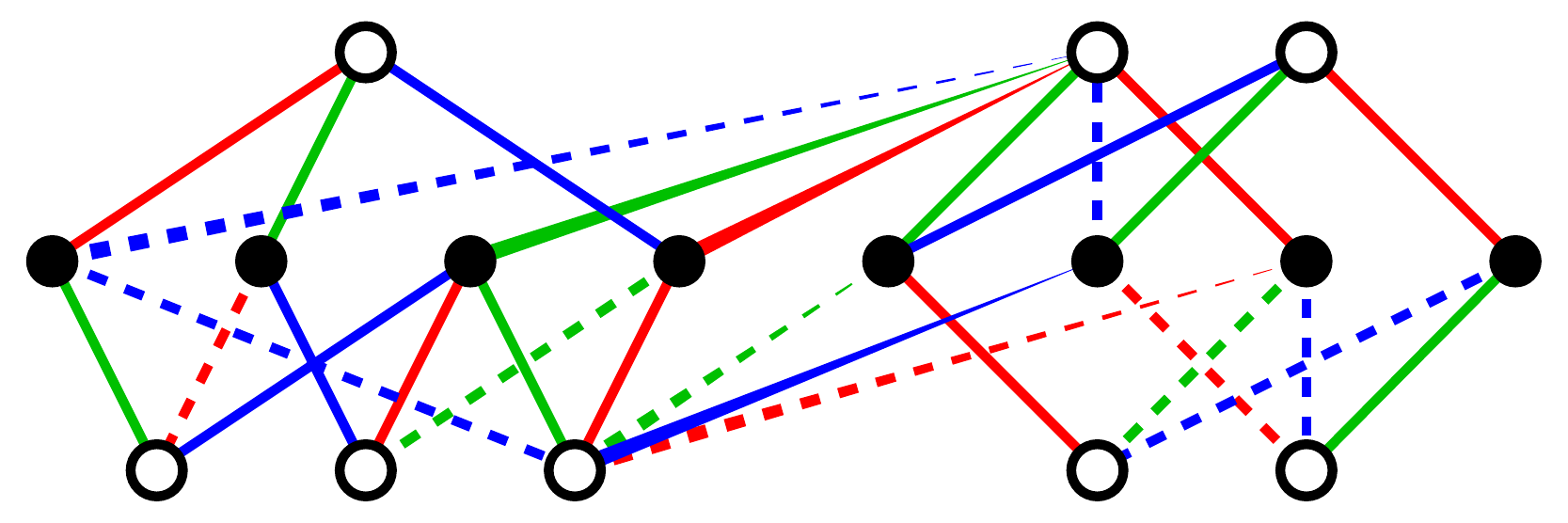}}
 \label{e:N3Qs}
\end{equation}
Reflecting the several steps in the procedure\eqs{e:L=-Y0}{e:QY-L3}, level by level and starting from the lowest one, the $(3|4|1)$ nodes in the Adinkra to the right of the ``$/$'' are used to gauge away 3, then 4 and finally one node from the direct sum of three $(1|3|3|1)$-Adinkras within the parentheses. The result on the right-hand side of the equality then clearly has no node in the bottom level, and has its $(5|8|3)$ nodes start at the next-to-lowest level.

\paragraph{Iterations:}
The foregoing suggest how to continue the iteration: the next step is depicted as
\begin{equation}
 \left.
  \left(\vC{\includegraphics[height=20mm]{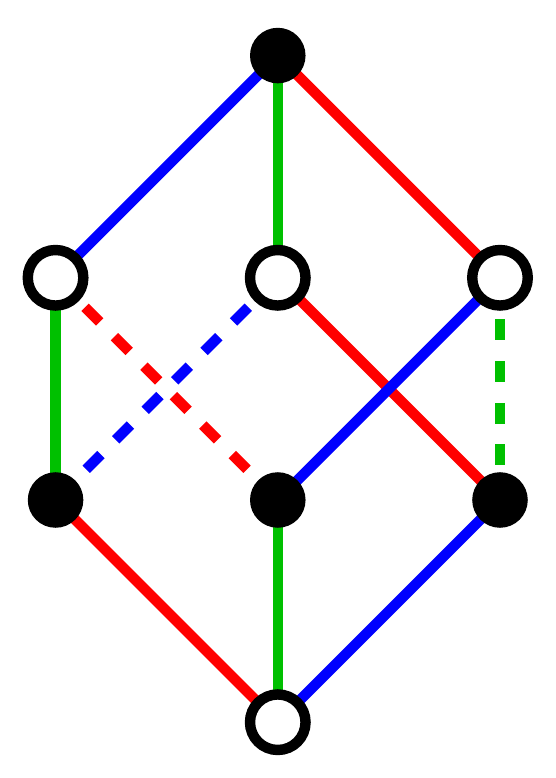}}
         \oplus
          \vC{\includegraphics[height=20mm]{N3B.pdf}}
           \oplus
            \vC{\includegraphics[height=20mm]{N3B.pdf}}
             \oplus
              \vC{\includegraphics[height=20mm]{N3B.pdf}}
               \oplus
                \vC{\includegraphics[height=20mm]{N3B.pdf}}\right)
  \right/\mkern-9mu
   \raisebox{-8.75mm}{\includegraphics[height=14mm]{YmodXn.pdf}}
 \label{e:N3Qn}
\end{equation}
the result of which will again have no node at the bottom level, and will have its $(7|12|5)$ nodes starting at the next-to-lowest level. The classification of Refs.\cite{r6-3c,r6-3,r6-3.2,r6-3.1} proves that this cannot be an Adinkra of $N\,{=}\,3$ worldline supersymmetry: it has more than $2^3=8$ nodes. Given the computations\eqs{e:L=-Y0}{e:QY-L3}, it is evident in turn that this next order gauge-quotient supermultiplet will again be depictable as a connected network of now three otherwise proper $N\,{=}\,3$ Adinkras---much as $\IY_I/(i\rD_I\IX)$ is a connected network of two otherwise proper $N\,{=}\,3$ Adinkras. 

This newly-minted $(7|12|5)$-component supermultiplet is then used to gauge away $7{+}12{+}5$ component fields from the direct sum of seven $(1|3|3|1)$-component supermultiplets leaving behind a $(0|9|16|7)$-component gauge-quotient supermultiplet depictable as a connected network of five otherwise proper $N\,{=}\,3$ Adinkras, and so on and so forth.

This iteration precisely matches the exact\footnote{See appendix~\ref{s:Smap} for a brief discussion and definition of exactness.} semi-infinite sequence~(8.28) in Ref.\cite{r6-1}:
\begin{equation}
 \vC{\begin{picture}(100,40)(30,6)
  \put(0,0){\includegraphics[width=5.5in]{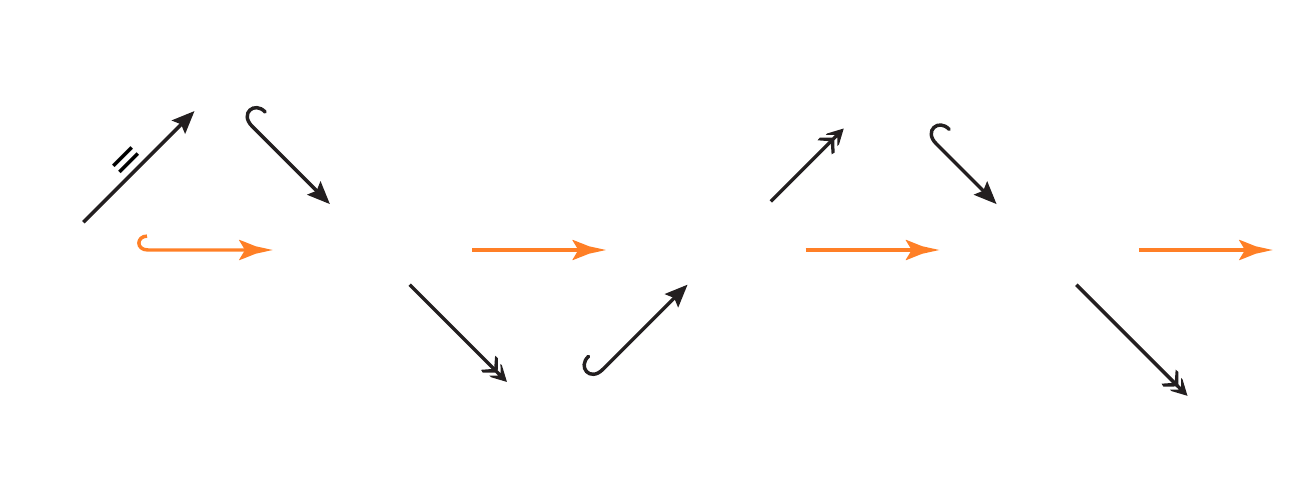}}
  \put(4.5,26){\large$\star$}
  \put(22,43){\large$\star$}
  \put(51,8){$(0|3|4|1)=\>$(\ref{e:QTDX})}
  \put(32,26){$(1|3|3|1)$}
  \put(68,26){$(3|9|9|3)$}
  \put(101,26){$(5|15|15|5)$}
  \put(87,42){$(0|5|8|3)=\>$(\ref{e:QY-L3})}
  \put(140,25.5){\etc}
  \put(128,8){\etc}
  \put(21,29){\C2{$\i_0$}}
  \put(55,29){\C2{$D^{\sss(1)}_I$}}
  \put(88,29){\C2{$\ID^{\sss(3)}_J{}^I$}}
  \put(126,29){\C2{$\ID^{\sss(5)}_K{}^J$}}
  \put(31,37){\small$\i_0$}
  \put(49,19){\small$\rD_I$}
  \put(64,17){\small$\i_1$}
  \put(78,35){\small$\ID_J{}^I$}
  \put(102,37){\small$\i_2$}
  \put(121,19){\small$\ID_K{}^J$}
 \end{picture}}
 \label{e:ZigZag}
\end{equation}
where ``$\star$'' represents the single constant (``zero-mode'') annihilated by the $\rD_I$-map, and the consecutive application of every two $\ID$-maps vanishes:
 $\ID^{\sss(3)}_J{}^I\circ D^{\sss(1)}_I=0$, 
 $\ID^{\sss(5)}_K{}^J\circ\ID^{\sss(3)}_J{}^I=0$, and so on; see Ref.\cite{r6-1} for details.
 Substituting the Adinkras and the Adinkra-like diagram from figure~\ref{f:A>B}, this is:
\begin{equation}
 \vC{\begin{picture}(100,44)(30,5)
  \put(0,0){\includegraphics[width=5.5in]{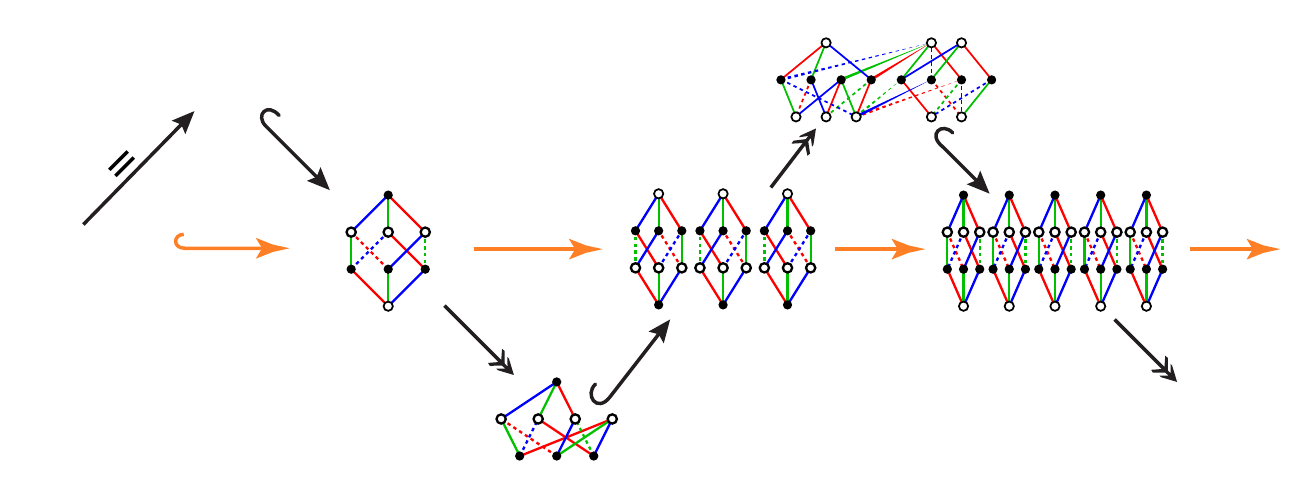}}
  \put(5.5,26){\large$\star$}
  \put(23,43){\large$\star$}
  \put(140,25.5){\etc}
  \put(126,8){\eq{e:N3Qn}}
  \put(21,29){\C2{$\i_0$}}
  \put(55,29){\C2{$D^{\sss(1)}_I$}}
  \put(90,29){\C2{$\ID^{\sss(3)}_J{}^I$}}
  \put(128,29){\C2{$\ID^{\sss(5)}_K{}^J$}}
  \put(32,38){\small$\i_0$}
  \put(50,19){\small$D^{\sss(1)}_I$}
  \put(64,17){\small$\i_1$}
  \put(76,35){\small$\ID^{\sss(2)}_J{}^I$}
  \put(104,37){\small$\i_2$}
  \put(123,17){\small$\ID^{\sss(3)}_K{}^J$}
  \put(108,44){=\,figure~\ref{f:A>B}}
 \end{picture}}
 \label{e:ZigZagA}
\end{equation}
Herein, all maps except $D^{\sss(1)}_I$ are supersymmetry-preserving strict homomorphisms of off-shell supermultiplets, so that the $Q$-action in their domain is properly mapped to the $Q$-action in their target. Also, the quotient\eq{e:N3Qn} fits as the target of the $\ID^{\sss(5)}_K{}^J$, the system of linear superderivative operators that has been precisely identified in Ref.\cite{r6-1}.

Furthermore, the double-barbed arrows (``$\onto$'') all denote {\em\/surjections\/}, which map to every nonzero element in their target some nonzero element from their domain; the hook-arrows (``$\into$'') all denote {\em\/injections\/}, which map every nonzero element of their domain to some nonzero element in their target. In fact, the latter maps are all supersymmetry-preserving inclusions, which are all defined akin to the procedure that resulted in table~\ref{t:dY=L}:
\begin{cons}\label{C:X>Y}
Given two supermultiplets represented by their superfield systems $\IM_1$ and $\IM_2$, the {\bsf\/cokernel\/} of the supersymmetry-preserving map $\IM_1\too{\m}\IM_2$ is constructed by mapping the like components, obtained by projecting like superderivatives to $\q^I\to0$
\begin{equation}
  ~\quad
  [\rD_{[I_1}\cdots \rD_{I_p]}]\big(\IM_2+\m(\IM_1)\big)\big|
  =[\rD_{[I_1}\cdots \rD_{I_p]}]\IM_2|+[\rD_{[I_1}\cdots \rD_{I_p]}]\m(\IM_1)|,
  \quad p=0\cdots N,
 \label{e:incl}
\end{equation}
and using component fields of $\m(\IM_1)$ to gauge away linear combinations of component fields of $\,\IM_2$ and is identified with the ``gauge-quotient,'' $\IM_2/\m(\IM_1)$. In turn, the {\bsf\/kernel\/} of $\m$ consists of elements of $\IM_1$ that are annihilated by the mapping $\IM_1\too{\m}\IM_2$.
\end{cons}
For the case\eq{e:N3Qs}, $\IM_2=\{\IY_1,\IY_2,\IY_3\}$, $\IM_1=\IX$ and $\m=i\rD_I$; notice that $\{Q_I,\m\}=0$.

The Reader familiar with the 4-dimensional supersymmetry literature will recognize that this provides a formal generalization of the two well-known cases of:
\begin{enumerate}\itemsep=-3pt\vspace{-2mm}
 \item The ``chiral superfield'' $\F$ is an example of the {\em\/kernel\/}-construction. The superderivative $\bar{\rD}_{\dt\a}$ maps and intact superfield to a spin-doublet of fermionic superfields: $\IE\too{\sss\bar{\rD}}\IF_{\dt\a}\Defl(\bar{\rD}_{\dt\a}\IE)$, and the chiral superfield is defined as the sub-supermultiplet of an intact superfield which is annihilated in this mapping, $\F\subset\IE:\bar{\rD}_{\dt\a}\F=0$.
 \item The ``vector superfield'' in the Wess-Zumino gauge is a representative of the {\em\/cokernel\/}-con\-struc\-tion. It is defined to consist of those components of the Hermitian superfield $\IV=\IV^\dag$ that remain after setting $\IV\simeq\IV+i(\bs\L-\bs\L^\dag)$, where $\bar{\rD}_{\dt\a}\bs\L=0$, and using components of $i(\bs\L-\bs\L^\dag)$ to gauge away (cancel, eliminate) components of $\IV$.
\end{enumerate}

The fact that all maps in the zig-zag sequences\eqs{e:ZigZag}{e:ZigZagA} are constructed using only simple inclusions\eq{e:incl} and superderivative operators guarantees that they preserve the $Q$-action since $\{Q_I,\rD_J\}=0$. By construction, the (increasing numbers of) intact supermultiplets inserted in the middle of these sequences are all off-shell. These two facts guarantee that the supermultiplets obtained in the peaks and valleys of these zig-zag sequences are also all off-shell.

\paragraph{Reduction:}
The one-way connectivity in the Adinkra-like graphical rendition in figure~\ref{f:A>B} implies that the supermultiplet $\IY_I/(i\rD_I\IX)$ can be reduced to a smaller supermultiplet, although it does not decompose into a direct sum of supermultiplets. (By contrast, if a unitary finite-dimensional representation of a classic Lie algebra reduces, it necessarily decomposes into a direct sum.)

Consider setting any one of the component fields in figure~\ref{f:A>B} to zero, say $\eY_{13}\to0$. Since $\C3{Q_1}(\eY_{13})=-\eU_1{}^2$ and $\C3{Q_1}(\eU_1{}^2)=-\eY_{13}$, the consistency of the $Q$-action implies that we must also set $\eU_1{}^2\to0$. Proceeding in this way, it is clear that we can set to zero only a complete $Q$-orbit, \ie, all component fields that can be reached one from another by $Q$-action.

For example, consider constraining all the left-hand side component fields in figure~\ref{f:A>B} to zero but not the right-hand side ones. Such a constraint is not compatible with the $Q$-action since, for example:
\begin{equation}
  \C3{Q_1}(\underbrace{\eY_{12}}_{\to0})
   = \underbrace{\eU_1{}^3}_{\to0} - \underbrace{\eU_3{}^1}_{\not\to0}
   \quad\To\quad
  \C3{Q_1}(0)\neq0,~~\text{inconsistent}.
\end{equation}
On the other hand, since supersymmetry does close on the right-hand half of the component fields in figure~\ref{f:A>B}, it {\em\/is\/} consistent to constrain
\begin{equation}
  \k:~~
  \eY_{22}=\eY_{23}=0=\EY_2=\EY_3,\quad
  \eU_2{}^1=\eU_2{}^2=\eU_2{}^3=\eU_3{}^1=0,
 \label{e:k}
\end{equation}
in the $(2|4|2)$-component right-hand half of the quotient as depicted in figure~\ref{f:A>B}. The $Q$-action of the $(3|4|1)$-component left-hand half then remains a complete supermultiplet, whereby $\IY_I/(i\rD_I\IX)$ has indeed been reduced (by constraining) to this smaller supermultiplet.

The asymmetry in the supermultiplet\eq{e:QY-L3}---the fact that the component fields\eq{e:k} {\em\/can\/} be constrained to zero consistently and independently whereas the complementary left-hand half of the component fields in figure~\ref{f:A>B} cannot---clearly stems from the fact that all $Q$-action across the partition in figure~\ref{f:A>B} is only one-way and only left-to-right. It is a general fact that if $C$ is the {\em\/extension\/} of the algebraic structure $B$ by $A$, so $A\into C\onto B$, then $C=A\oplus B$ if and only if the surjection $C\onto B$ has an inverse and $A\into C\>\reflectbox{$\into$}\>B$, whereby both $A$ and $B$ are proper algebraic sub-structures of $C$.
The fact that the right-hand half of the supermultiplet in figure~\ref{f:A>B} is {\em\/not\/} a proper sub-supermultiplet\eq{e:QY-L3} then implies that the supermultiplet\eq{e:QY-L3} is not so decomposable.

In fact, the constraint\eq{e:k} then provides an inclusion of the $(3|4|1)$-component sub-super\-mul\-ti\-plet in the $(5|8|3)$-component gauge-quotient\eq{e:N3Qs}, and defines
\begin{equation}
 \vC{\begin{picture}(120,20)
   \put(0,0){\includegraphics[height=20mm]{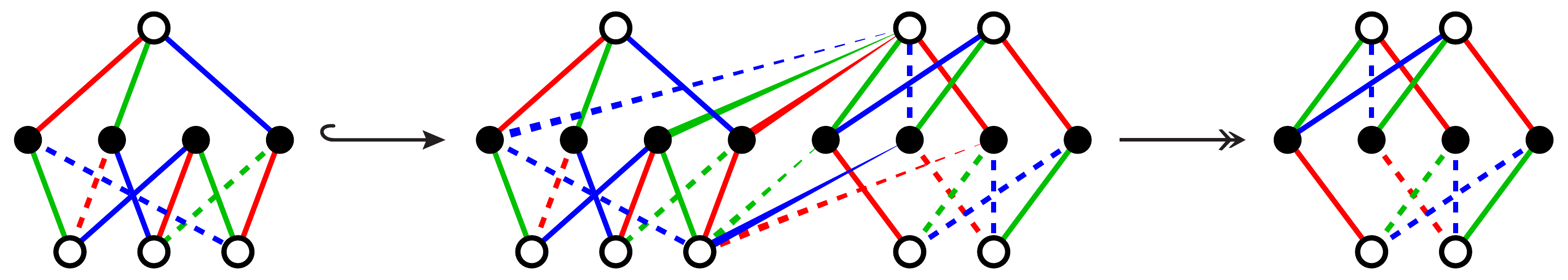}}
    \put(2,15){$\IA$}
    \put(108,15){$\IB$}
    \put(26,12){$\k$}
    \put(23,6){\small(\ref{e:k})}
 \end{picture}}
 \label{e:A>AB>B}
\end{equation}
which also proves that $[\IY_I/(i\rD_I\IX)]/\IA=\IB$:
\begin{equation}
 \vC{\begin{picture}(120,25)
   \put(0,11){$\left(\vC{\includegraphics[height=20mm,viewport=260 0 640 160,clip]
            {YmodXx.pdf}}\right)$}
   \put(55,11){\large$\Bigg/$}
   \put(60,2){\includegraphics[height=20mm,viewport=0 0 180 160,clip]
            {YmodXx.pdf}}
   \put(85,11){\large$=$}
   \put(90,2){\includegraphics[height=20mm,viewport=720 0 900 160,clip]
            {YmodXx.pdf}}
 \end{picture}}
 \label{e:AB/A}
\end{equation}
so that the supermultiplet\eq{e:QY-L3} may itself be both
 constrained\eq{e:k} so as to reduce to a sub-supermultiplet shown at the far left of\eq{e:A>AB>B}, as well as
 further gauge-quotiented\eq{e:AB/A}. One says that the $(5|8|3)$-component supermultiplet\eq{e:QY-L3} is an {\em\/extension\/} of the $(2|4|2)$-component Adinkra supermultiplet $\IB$ by the $(3|4|1)$-component Adinkra supermultiplet $\IA$. (Having spelled out the details of such a quotient above\eqs{e:QTY}{e:QY-L3}, the Adinkra display\eq{e:AB/A} should suffice here.)

\paragraph{Main result:}
Whereas we have in the foregoing analyzed in detail only the particular gauge-quotient\eq{e:N3Qs}, the zig-zag sequence\eq{e:ZigZagA} makes it clear that this is but the first non-Adinkra supermultiplet appearing in this sequence. It is fairly evident that in fact all supermultiplets subsequently constructed in this sequence (and that appear in the zig-zag peaks and valleys) are not proper Adinkras and so are not limited by the classification theorems of Refs.\cite{r6--1,r6-3,r6-3.2,r6-3.1}. Furthermore, whereas Ref.\cite{r6-1} explicitly shows that in the analogous sequence for $N\,{=}\,2$ all supermultiplets {\em\/are\/} Adinkras, it is clear that the supermultiplets constructed by the analogous sequences for all $N\,{>}\,2$ will not be Adinkras, just as the one in figure~\ref{f:A>B} is not. We thus conclude:
\begin{thrm}[Main Result]\label{T:MR}
The semi-infinite sequence of direct sums of an indefinitely growing number of intact supermultiplets (Adinkras $\,\IE_k$), such as the horizontal sequence in\eq{e:ZigZagA}, may be resolved into a zig-zag sequence as shown in\eq{e:ZigZagA}.
 This zigzag sequence then constructs the supermultiplets that appear in the zig-zag peaks and valleys, \ie, the end-points in the ``$\IA_k\overset{\m_k}{\into}\oplus_k\IE_k\onto\IB_k$''-form three-part segments, so-called ``short exact sequences,'' and where the maps $\m_k$ are defined as in construction~\ref{C:X>Y}.
 
 More explicitly, this defines a sequence of supermultiplets:
\begin{equation}
 \IA_k\Defl\big\{\oplus_k\IE_k\colon\m_k(\oplus_k\IE_k)=0\big\}
  \quad\text{and}\quad
 \IB_k\Defl\big\{\oplus_k\IE_k\simeq\oplus_k\IE_k+\m_k(\oplus_k\IE_k)\big\}
 \label{e:AkBk}
\end{equation}
\begin{enumerate}\itemsep=-3pt\vspace*{-5mm}\addtolength{\leftskip}{2pc}
 \item indefinite in number (\,$k=1,3,5,\dots$),
 \item indefinitely growing in size (\,=\,number of component fields, \ie, nodes),
 \item depictable as connected networks of Adinkras.
\end{enumerate}
\end{thrm}
Alternatively, this says that direct {\em\/sums\/} of intact Adinkras may be reduced to supermultiplets that are depicted by connected networks of Adinkras that do not themselves decompose into a direct sum of Adinkras---which are certainly not intact. This result is in {\em\/contradistinction\/} to the corresponding situation in the representation theory of classic Lie algebras, where direct sums of irreducible representations do not reduce in any other way.

\section{Conclusions}
\label{s:Coda}
Adinkras depict supermultiplets of $N$-extended worldline supersymmetry without central charges and are restricted to have the (chromo-)topology of $N$-cubes or their $k$-fold quotients by certain $\ZZ_2$-reflections\cite{r6-3.1}. In particular, this means that all such supermultiplets have $2^{N-k}$ component fields: one half bosons, the other half fermions.

Herein, we have shown that direct sums of intact Adinkras (figure~\ref{f:Yk}) may be reduced by means of gauging away another Adinkra (figure~\ref{f:DX}), producing supermultiplets (figure~\ref{f:A>B}) that:
\begin{enumerate}\itemsep=-3pt\vspace{-3mm}
 \item do not decompose into a direct sum of Adinkras;
 \item are themselves not depictable by a proper Adinkra\cite{r6-3.1};\newline see also Refs.\cite{rA,r6-1,r6--1,r6-3c,r6-3,r6-3.2,r6-1.2,r6-3.4};
 \item are not limited in size: see\eq{e:N3Qn}, the sequence\eq{e:ZigZagA}, and theorem~\ref{T:MR};
 \item are depicted as connected networks of otherwise proper Adinkras\eq{e:A>AB>B}.
\end{enumerate}\vspace{-3mm}
While we have examined in detail the construction that starts with directs sums of a growing number of intact Adinkras, it is clear that the construction is straightforward to generalize so as to use all other types of Adinkras instead of the intact ones. The precise conditions under which this can be done are however beyond our present scope.

In the constructions leading to theorem~\ref{T:MR} and in the developing a comprehensive off-shell representation theory of supersymmetry,
 Adinkras play the role somewhat akin to the role that single boxes within Young tableaux play in depicting the fundamental representation in Lie algebras; see table~\ref{t:BoxDinkra}.
\begin{table}[htb]
\centering
  \begin{tabular}{@{} >{\raggedleft\baselineskip=11pt\bfseries}p{42mm}|
                      >{\raggedright\baselineskip=11pt}p{45mm}|
                      >{\raggedright\baselineskip=11pt}p{59mm} @{}p{0pt}@{}}
 & {\bf Lie algebra irrreps (Weyl construction)}
 & {\bf Off-shell supermultiplets} &\\
    \midrule[.667pt]
 Starting object(s) \goodbreak and their depiction
                     & fundamental irrep (\,$\Box$\,)
                     & Adinkras ($\vC{\includegraphics[height=10mm]{N3B.pdf}}$
                             \& $\vC{\includegraphics[height=10mm]{N3T.pdf}}$
                             \& $\cdots$) &\\
 \midrule
 Combining operator  & ~~$\otimes$ & ~~$\oplus$ &\\ 
 \midrule
 Reduction methods   & Young symmetrization, traces w/inv. tensors
                     & Construction~\ref{C:X>Y}:
                       $\ker(\m)$ and $\cok(\m)$ of supersymmetric maps &\\ 
 \midrule
 Resulting objects\goodbreak and their depiction
                     & arbitrarily large irreps,
                       \& their Young tableaux (1-quadrant graphs)
                     & networks of otherwise proper Adinkras,
                       connected by\goodbreak one-way $Q$-action edges
                        &\\ 
    \bottomrule
  \end{tabular}
 \caption{The conceptual relation between boxes and Young tableaux depicting irreducible representations of classic Lie algebras, {\em\/vs\/}.\ Adinkras and their connected networks used herein to depict indecomposable off-sell representations of $N$-extended worldline supersymmetry without central charges.}
 \label{t:BoxDinkra}
\end{table}
Just as single boxes are put together according to specific rules and result in Young tableaux that depict irreducible representations of classic Lie algebras, Adinkras can be connected by additional one-way edges (depicting one-way $Q$-action) into networks that depict indecomposable representations of the $N$-extended supersymmetry algebra on the worldline. While figure~\ref{f:A>B} provides but the simplest such example, the sequence\eq{e:ZigZagA} provides an indefinite sequence of ever larger such examples.

In fact, the analogy with the construction of representations of classical Lie algebras can be made even more obvious, by noting that, for example:
\begin{equation}
 \mathfrak{su}(3):\qquad
 {\bf3}\,{\otimes}\,{\bf3}\ominus{\bf3^*} = {\bf6},\quad
 {\bf3}\,{\otimes}\,{\bf3^*}\ominus{\bf1} = {\bf8},\quad\etc
 \label{e:LieR}
\end{equation}
construct the representations $\bf6$ (reps.\ $\bf8$) from a direct product of $\bf3$ and $\bf3$ (reps.\ $\bf3$ and $\bf3^*$) by {\em\/subtracting\/} a $\bf3^*$ (reps.\ $\bf1$). Instead of this subtraction (which is not possible in supersymmetry), Construction~\ref{C:X>Y} ``subtracts'' degrees of freedom by constructing a quotient with respect to a supersymmetric map\eq{e:Y/iDX}. Also, Construction~\ref{C:X>Y} uses $\oplus$ in place of the $\otimes$ in the equations\eq{e:LieR}.

As shown in\eqs{e:k}{e:AB/A}, the supermultiplet depicted in figure~\ref{f:A>B} does not {\em\/decompose\/}, but it does {\em\/reduce\/}: one ``half'' of it may be isolated by constraining\eq{e:k}, the other ``half'' by gauging\eq{e:AB/A}, but the supermultiplet cannot be decomposed into a direct sum of two independent supermultiplets. We expect such reductions in larger supermultiplets of this kind, such as\eq{e:N3Qn}, to also be possible, but be more and more complicated.

\bigskip
Finally, on dimensional grounds, a nontrivial Lagrangian---{\em\/if it exists\/}---for the supermultiplet depicted in figure~\ref{f:A>B} and using the basis\eq{e:Y/iDX2} may be written in the form
\begin{equation}
 \begin{aligned}
 L&= A^{k\ell,IJ}\,\Dt{\eY}_{kI}\,\Dt\eY_{\ell J}
    +A^{k\ell,I}\,\Dt{\eY}_{kI}\,\EY_\ell
    +A^{k\ell}_{IJ}\,\Dt{\eU}_k{}^I\,{\eU}_\ell{}^J
    +A^{k\ell}\,\EY_k\,\EY_\ell\\
 &\mkern20mu
 +B^{k\ell,IJ}\,\eY_{kI}\,\Dt\eY_{\ell J}
 +B^{k\ell,I}\,\eY_{kI}\,\EY_\ell
 +B^{k\ell}_{IJ}\,{\eU}_k{}^I\,{\eU}_\ell{}^J
 ~+~C^{k\ell,IJ}\,\eY_{kI}\,\eY_{\ell J}
 ~+~D^k\,\EY_k.
 \end{aligned}
 \label{e:L}
\end{equation}
The $A$,- $B$- and $C$-coefficients are determined by requiring that $\e^IQ_I(L)=\e^I\Dt{K}_I$, where $K_I$ are some functional expressions of the component fields\eq{e:Y/iDX2}. In principle, this requirement may reduce the nonlinear terms in\eq{e:L} to a total time-derivative. The verification that nontrivial choices of coefficients in\eq{e:L} do exist, their concrete choice and interpretation in any particular model is however outside our present scope. In turn, the $D$-coefficients remain arbitrary since $Q_I(\EY_k)$ are all total $\t$-derivatives; see\eq{e:QY-L3}.

We reiterate that out focus on $N\,{=}\,3$ worldline supersymmetry is solely for simplicity. This disentangles the basic features of the construction as itemized and summarized above, from the added technical detail and complexity stemming from additional symmetries, such as gauge and Lorentz symmetries, and additional (complex, hyper-complex and other) structures.

\paragraph{\bfseries Acknowledgments:}
 TH is grateful to the Department of Energy for the generous support through the grant DE-FG02-94ER-40854, as well as the Department of Physics, University of Central Florida, Orlando FL, and the Physics Department of the Faculty of Natural Sciences of the University of Novi Sad, Serbia, for recurring hospitality and resources.

\appendix
\section{Three Supermultiplets}
 \label{s:X+DX}
\paragraph{The Bosonic Intact Supermultiplet:}
The component fields of this $(1|3)$-superfield, $\IX$, are {\em\/defined\/} by means of superderivative projections\cite{r1001,rBK}
\begin{subequations}
 \label{e:cX}%
\begin{alignat}9
 x     &\Defl \IX|,&\qquad\qquad
 \x_I  &\Defl i\rD_I\IX|,\\*
 X^I   &\Defl \frc{i}{2!}\ve^{IJK}\,\rD_{[J}\rD_{K]}\IX|,&\qquad\qquad
 \X    &\Defl-\frc1{3!}\ve^{IJK}\,\rD_{[I}\rD_J\rD_{K]}\IX|,
\end{alignat}
\end{subequations}
where `$|$' indicates projection to the (bosonic) worldline subspace of the $(1|3)$-superspace, \ie, setting the (fermionic) superspace coordinates $\q^I$ to zero and the numerical coefficients are chosen for later convenience. A bosonic intact supermultiplet thus consists of the component fields
\begin{equation}
  \IX = (x\mid \x_1,\x_2,\x_3\mid X^1,X^2,X^3\mid \X)
\end{equation}
and we refer to it as $(1|3|3|1)$-dimensional representation of the worldline 3-supersymmetry. The indicated basis of component fields is graded by their relative engineering dimension
\begin{equation}
  [\x_I]=[x]+\inv2,\quad
  [X^I]=[\x_I]+\inv2=[x]+1,\quad
  [\X]=[X^I]+\inv2=[\x_I]+1=[x]+\frc32.
 \label{e:EDX}
\end{equation}
Using\eq{e:Q=DFJ}, we compute:
\begin{subequations}
 \label{e:QcX}%
\begin{alignat}9
 Q_I(x)&=i\rD_I(\IX)\big|=\x_I,\\[2mm]
 Q_I(\x_J)&=-i\rD_I(i\rD_J\IX)\big|
  =\big[i\d_{IJ}\vdt+\rD_{[I}\rD_{J]}\big]\IX\big|
  =i\d_{IJ}\Dt{x}-i\ve_{IJK}\,X^K,\\[2mm]
 Q_I(X^J)&=iD_I(\frc{i}{2!}\ve^{JKL}\,\rD_{[K}\rD_{L]}\IX)\big|
  =-\ve^{JKL}[i\d_{I[K}D_{L]}\vdt+\frc12D_{[I}D_KD_{L]}]\IX\big|
  =\ve_I{}^{JK}\Dt\x_K+\d_I{}^J\,\X,\\[2mm]
 Q_I(\X)&=-i\rD_I(-\frc1{3!}\ve^{JKL}\rD_J\rD_K\rD_L\IX)\big|
  =\frc{i}{3!}\ve^{JKL}\big[0+3i\d_{I[J}\rD_K\rD_{L]}\vdt\big]\IX\big|
  =i\d_{IJ}\,\Dt{X}^J,
\end{alignat}
\end{subequations}
and summarize this in tabular form:
\begin{equation}
  \begin{array}{@{} c@{:~~}c|ccc|ccc|c @{}}
 \omit& x & \x_1 & \x_2 & \x_3 & X^1 & X^2 & X^3 & \X \\[1pt]
    \toprule
\C3{Q_1} & \x_1 & i\Dt{x} & -iX^3   & iX^2    & \X       & \Dt\x_3  & -\Dt\x_2 & i\Dt{X}^1 \\ 
\C1{Q_2} & \x_2 & iX^3    & i\Dt{x} & -iX^1   & -\Dt\x_3 & \X       & \Dt\x_1  & i\Dt{X}^2 \\ 
\C6{Q_3} & \x_3 & -iX^2   & iX^1    & i\Dt{x} & \Dt\x_2  & -\Dt\x_1 & \X       & i\Dt{X}^3 \\ 
    \bottomrule
  \end{array}
 \label{e:QTX}
\end{equation}
This makes it evident that in $\IX$, each $Q_I$-transformation of each of the component fields consists of a constant multiple of precisely one other component superfield or its $\vdt$-derivative and so can be depicted by the Adinkra\cite{r6-1}:
\begin{equation}
 \vC{\begin{picture}(160,38)
   \put(.5,0){\includegraphics[width=160mm]{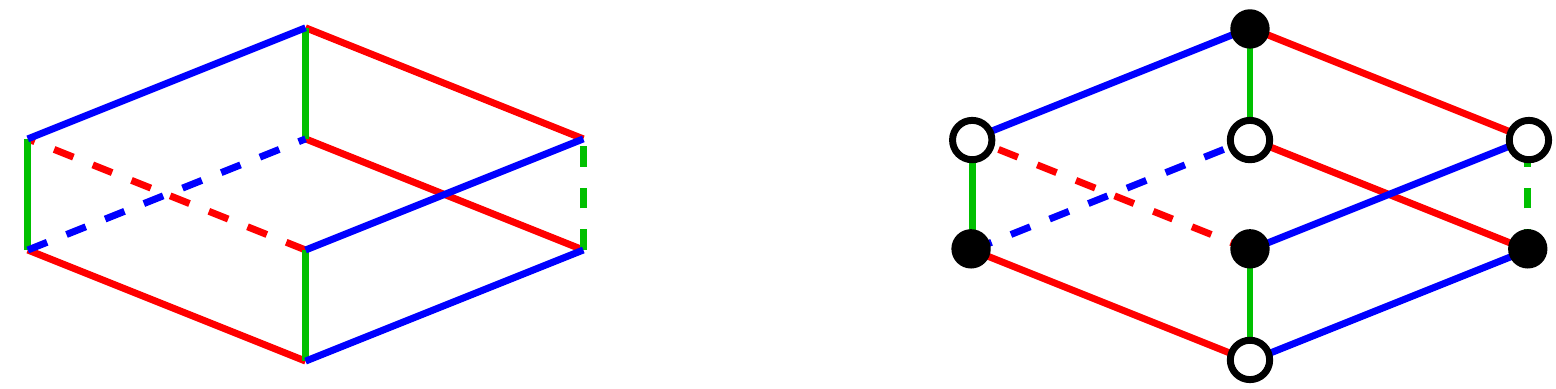}}
     \put(32,36){\cB{$\X$}}
     \put(3,24){\cB{$X^3$}}
     \put(32,24){\cB{$X^2$}}
     \put(61,24){\cB{$X^1$}}
     \put(3,13){\cB{$\x_1$}}
     \put(32,13){\cB{$\x_2$}}
     \put(61,13){\cB{$\x_3$}}
     \put(32,2){\cB{$x$}}
     \put(74,20){or simply}
 \end{picture}}
 \label{e:AX}
\end{equation}
where the dashed edges encode the negative signs in\eq{e:QTX}, and the nodes are drawn at heights proportional to the engineering dimension\eq{e:EDX} of the component fields that they depict.

\paragraph{The Fermionic Intact Supermultiplet:}
The component fields of the $k^\text{th}$ of such $(1|3)$-superfields, $\IY_k$, are {\em\/defined\/} by the superderivative projections\cite{r1001,rBK}:
\begin{subequations}
 \label{e:cY}%
\begin{alignat}9
 \h_k       &\Defl \IY_k|,&\qquad\qquad
 Y_{kJ}     &\Defl -\rD_J\IY_k|,\\
 \Y_k{}^I   &\Defl -\frc{i}{2!}\ve^{IJK}\rD_{[J}\rD_{K]}\IY_k|,&\qquad\qquad
 {\cal Y}_k &\Defl \frc{i}{3!}\ve^{JKL}\rD_{[J}\rD_K\rD_{L]}\IY_k|.
\end{alignat}
\end{subequations}
A fermionic intact supermultiplet thus consists of the component fields
\begin{equation}
  \IY_k = (\h_k\mid Y_{k1},Y_{k2},Y_{k3}\mid \Y_k{}^1,\Y_k{}^2,\Y_k{}^3\mid {\cal Y}_k),
\end{equation}
where
\begin{equation}
  [Y_{kJ}]=[\h_k]+\inv2,\quad
  [\Y_k{}^J]=[Y_{kJ}]+\inv2=[\h_k]+1,\quad
  [{\cal Y}_k]=[\Y_k{}^J]+\inv2=[Y_{kJ}]+1=[\h_k]+\frc32.
 \label{e:EDY}
\end{equation}

For the components\eq{e:cX} of the intact $(1|3)$-supermultiplet we compute:
\begin{subequations}
 \label{e:QcY}%
\begin{alignat}9
 Q_J(\h_k)&=-i\rD_J(\IY_k)\big|=i(-\rD_J\IY_k)\big|
 =iY_{kJ},\\[2mm]
 Q_J(Y_{kK})&=i\rD_J(-\rD_K\IY_k)\big|
  =-i\big[i\d_{JK}\vdt+\rD_{[J}\rD_{K]}\big]\IY_k\big|
 =\d_{JK}\Dt\h_k+\Y_k{}^L\ve_{JKL},\\[2mm]
 Q_J(\Y_k{}^K)&=-iD_I(\frc{i}{2!}\ve^{KLM}\,\rD_{[L}\rD_{M]}\IY_k)\big|
  =\ve^{KLM}[i\d_{I[L}D_{M]}\vdt+\frc12D_{[I}D_LD_{M]}]\IY_k\big|,\nn\\*
  &~~~=-i\,\ve_J{}^{KL}\Dt{Y}_{kL}+i\d^K\!_J\,{\cal Y}_k,\\[2mm]
 Q_J({\cal Y}_k)&=i\rD_J(\frc{i}{3!}\ve^{KLM}\rD_K\rD_L\rD_M\IY_k)\big|
  =-\frc1{3!}\ve^{KLM}\big[0+3i\d_{J[K}\rD_L\rD_{M]}\vdt\big]\IY_k\big|
  =\d_{JK}\Dt{\Y}_k{}^K,
\end{alignat}
\end{subequations}
which are summarized in a tabular manner:
\begin{equation}
  \begin{array}{@{} c@{:~~}c|ccc|ccc|c @{}}
 \omit   & \h_k    & Y_{k1}    & Y_{k2}    & Y_{k3}
          & \Y_k{}^1 & \Y_k{}^2 & \Y_k{}^3 & \cY_k \\[1pt]
    \toprule
\C3{Q_1} & iY_{k1} & \Dt\h_k   & \Y_k{}^3  & -\Y_k{}^2
          & i\cY_k       & -i\Dt{Y}_{k3}  & i\Dt{Y}_{k2} & \Dt\Y_k{}^1 \\ 
\C1{Q_2} & iY_{k2} & -\Y_k{}^3 & \Dt\h_k   & \Y_k{}^1
          & i\Dt{Y}_{k3}  & i\cY_k        & -i\Dt{Y}_{k1}  & \Dt\Y_k{}^2 \\ 
\C6{Q_3} & iY_{k3} & \Y_k{}^2  & -\Y_k{}^1 & \Dt\h_k
          & -i\Dt{Y}_{k2} & i\Dt{Y}_{k1}   & i\cY_k   & \Dt\Y_k{}^3 \\ 
    \bottomrule
  \end{array}
 \label{e:QTYa}
\end{equation}
and so may be depicted by the Adinkra:
\begin{equation}
 \vC{\begin{picture}(160,40)
   \put(.5,0){\includegraphics[width=160mm]{N3IntactF.pdf}}
     \put(32,36){\cB{$\cY_k$}}
     \put(3,24){\cB{$\Y_k{}^3$}}
     \put(32,24){\cB{$\Y_k{}^2$}}
     \put(61,24){\cB{$\Y_k{}^1$}}
     \put(3,13){\cB{$Y_{k1}$}}
     \put(32,13){\cB{$Y_{k2}$}}
     \put(61,13){\cB{$Y_{k3}$}}
     \put(32,2){\cB{$\h_k$}}
     \put(74,20){or simply}
 \end{picture}}
 \label{e:AYa}
\end{equation}
Notice that the fermionic intact Adinkra\eq{e:AY} equals the bosonic intact Adinkra\eq{e:AX}, drawn however upside-down. Equivalently, one can obtain\eq{e:AY} by re-drawing\eq{e:AX} with the statistics/color of the nodes flipped (white\,$\iff$\,black) and flipping the sign of the upper half of the nodes.

\paragraph{A superderivative superfield:}
Given the original component field definitions\eq{e:cX}, we easily compute the component field expressions of the superdifferential ``gradient'' $(i\rD_I\IX)$:
\begin{subequations}
\begin{alignat}9
 (i\rD_I\IX)|&=\x_I,\label{e:DX0}\\
 i\rD_J(i\rD_I\IX)|
 &=-\big[i\d_{IJ}\vdt-\rD_{[I}\rD_{J]}\big]\IX|
  = -i\d_{IJ}\Dt{x}-i\ve_{IJK}X^K,\label{e:DX1}\\
 i\rD_{[J}\rD_{K]}(i\rD_I\IX)|
 &=-\big[2i\rD_{[J}\d_{K]I}\vdt+\rD_{[I}\rD_J\rD_{K]}\big]\IX|
  =2\d_{I[J}\Dt\x_{K]}+\ve_{IJK}\,\X,\label{e:DX2}\\
 \rD_{[J}\rD_K\rD_{L]}(i\rD_I\IX)|
 &=i\big[3i\rD_{[J}\rD_K\d_{L]I}\vdt\big]\IX|
  =\ve_{JKL}\,\d_{IM}\Dt{X}^M.\label{e:DX3}
\end{alignat}
\end{subequations}
Notice that $\x_I$ occurs at lowest level\eq{e:DX0}, and that only the $\t$-derivative of $x$ occurs and at the second level\eq{e:DX1}. We have thus defined a supermultiplet
\begin{equation}
  \IL_I=(\l_I\mid L^0,L^1,L^2,L^3\mid \L),
 \label{e:Y}
\end{equation}
for which the identification $\IL_I=(i\rD_I\IX)$ induces the component-level identifications
\begin{subequations}
 \label{e:cL=cDX}%
\begin{alignat}9
 \l_I&=(i\rD_I\IX)|&&
     &&=\x_I,\\
 L^0 &=\frc{i}3\d^{IJ}\big(i\rD_I(i\rD_J\IX)\big)\big|&
     &=-\frc{i}3\Big(\big[{\ttt\sum_{I=1}^3}i\vdt\big]\IX\Big)\Big|
     &&=\Dt{x},\\
 L^I &=\inv{2i}\e^{IJK}\big(i\rD_J(i\rD_K\IX)\big)|&
     &=\inv{2!}\e^{IJK}(i\rD_J\rD_K\IX)|
     &&=X^I,\\
 \L  &=\frc1{3!}\e^{IJK}\big(i\rD_I\rD_J(i\rD_K\IX)\big)|&
     &=\frc1{3!}\e^{IJK}(-\rD_I\rD_J\rD_K\IX)|
     &&=\X.
\end{alignat}
\end{subequations}
Owing to the relation $\IL_I=(i\rD_I\IX)$, the supermultiplet satisfies the superderivative constraints:
\begin{equation}
 \rD_1\IL_1=\rD_2\IL_2=\rD_3\IL_3,
  \quad\text{and}\quad
 \rD_I\IL_J=-\rD_J\IL_I,~~I\neq J.
 \label{e:conIL}
\end{equation}
In turn, the superderivative constraints\eq{e:conIL} may be {\em\/solved\/} by setting $\IL_I=(i\rD_I\IX)$.

This relationship allows reading off the supersymmetry transformation rules for $\IL_I$ from those for the bosonic intact supermultiplet\eq{e:QcX}. In tabular form, akin to Table\eq{e:QTX}, these are:
\begin{equation}
  \begin{array}{@{} c@{:~~}ccc|cccc|c @{}}
 \omit& \l_1 & \l_2 & \l_3 & L^0 & L^1 & L^2 & L^3 & \L \\[1pt] 
    \toprule
\C3{Q_1} &  iL^0 & -iL^3 &  iL^2 & \Dt\l_1 &  \L      &  \Dt\l_3 & -\Dt\l_2 & i\Dt{L}^1 \\ 
\C1{Q_2} &  iL^3 &  iL^0 & -iL^1 & \Dt\l_2 & -\Dt\l_3 &  \L      &  \Dt\l_1 & i\Dt{L}^2 \\ 
\C6{Q_3} & -iL^2 &  iL^1 &  iL^0 & \Dt\l_3 &  \Dt\l_2 & -\Dt\l_1 &  \L      & i\Dt{L}^3 \\ 
    \bottomrule
  \end{array}
 \label{e:QDT}
\end{equation}
and are depicted by the Adinkra
\begin{equation}
 \vC{\begin{picture}(160,42)
   \put(.5,0){\includegraphics[width=160mm]{N3D.pdf}}
     \put(32,39){\cB{$\L$}}
     \put(3,21){\cB{$L^3$}}
     \put(22,21){\cB{$L^2$}}
     \put(40,21){\cB{$L^1$}}
     \put(58,21){\cB{$L^0$}}
     \put(13,2){\cB{$\l_1$}}
     \put(31,2){\cB{$\l_2$}}
     \put(50,2){\cB{$\l_3$}}
     \put(74,20){or simply}
 \end{picture}}
 \label{e:AY}
\end{equation}

\section{Supersymmetric Mapping}
\label{s:Smap}
The relation $\IL_I=(i\rD_I\IX)$ in fact provides a mapping from the bosonic intact supermultiplet $\IX$ to the
constrained\eq{e:conIL} supermultiplet triple $(\IL_1,\IL_2,\IL_3)$:
\begin{equation}
  \IX \too{~i\rD_I~} \IL_I\big|_{\text{(\ref{e:conIL})}}.
\end{equation}
The identifications\eq{e:cL=cDX} make this mapping explicit at the level of component fields and show that this mapping is {\em\/almost\/} 1--1: all the component fields are identified exactly, except for $L^0=\Dt{x}$, which omits the constant mode in any power-expansion of $x=x(\t)$. Therefore,
\begin{equation}
 \ker\big(\IX \too{~i\rD_I~} \IL_I\big) = x(0),
\end{equation}
and the sequence of maps (with $\i$ denoting the simple identification map)
\begin{equation}
  0 \to x(0) \too{~\i~}\IX \too{~i\rD_I~} \IL_I,
   \qquad\text{or}\qquad
  x(0)\overset{\i}{\into}\IX \too{~i\rD_I~} \IL_I,
 \label{e:L=iDX}
\end{equation}
is said to be {\em\bsf\/exact\/}: the kernel of every map is the image of the preceding map. Indeed, the map $\i$ identifies $x(0)$ with the constant mode in a power expansion of $x(\t)$. Then,
\begin{enumerate}\itemsep=-3pt\vspace{-2mm}
 \item $\text{image}\big(0\to x(0)\big)=0=\ker(\i)$;
 \item $\text{image}\big(x(0)\too{\i}\IX\big)=x(0)=\ker(iD_I)$.
\end{enumerate}\vspace*{-2mm}
In particular, $i\rD_I\circ\i\id0$. Equally importantly, the maps (anti)commute with the supersymmetry generators, $[Q_I,\i]=0=\{Q_I,i\rD_J\}$, and so are {\em\bsf\/supersymmetry-preserving\/}.
 One then says that the whole sequence of maps\eq{e:L=iDX} is {\em\bsf\/supersymmetry-equivariant\/}.
 It is this {\em\/exactness\/} of the sequence of {\em\/supersymmetry-preserving\/} maps and the fact that the maps satisfy the definition~\ref{linalg} that permits defining the off-shell supermultiplet $\IL_I$ in terms of the off-shell supermultiplet $\IX$.

This same exactness of the sequence of supersymmetry-preserving and definition~\ref{linalg}-abiding maps also holds throughout the semi-indefinite zig-zag sequence\eqs{e:ZigZag}{e:ZigZagA}. This guarantees that the indefinitely many supermultiplets defined in its peaks and valleys of\eqs{e:ZigZag}{e:ZigZagA} are all proper off-shell supermultiplets because those in the horizontal sequence (the direct sums of increasingly more intact supermultiplets) are.

\small\raggedright

\end{document}